\documentclass{aa}  
\usepackage{graphicx}
\usepackage{txfonts}
\usepackage{natbib}
\bibpunct{(}{)}{;}{a}{}{,} 
\usepackage{amssymb}
\usepackage{longtable}
\usepackage[breaklinks=true]{hyperref}
\hypersetup{colorlinks,linkcolor=red,citecolor=blue,urlcolor=blue}
\usepackage[backgroundcolor=white]{todonotes}
\newcommand{\Teff}{$T_{\rm eff}$}
\newcommand{\logg}{$\log g$}
\newcommand{\FeH}{[Fe/H]}

\begin{document}
   \title{On the metallicity of open clusters}

   \subtitle{II. Spectroscopy}

   \author{U. Heiter
          \inst{\ref{inst:Uppsala}}
          \and
          C. Soubiran
          \inst{\ref{inst:Bordeaux}}
          \and
          M. Netopil
          \inst{\ref{inst:Wien}}
          \and
          E. Paunzen
          \inst{\ref{inst:Brno}}
          }

   \institute{
              Institutionen f\"or fysik och astronomi, Uppsala universitet,
              Box 516, 751\,20 Uppsala, Sweden\\
              \email{ulrike.heiter@physics.uu.se}
              \label{inst:Uppsala}
         \and
              Universit\'e Bordeaux 1, CNRS, Laboratoire d’Astrophysique de Bordeaux UMR 5804,
              BP 89, 33270 Floirac, France
              \label{inst:Bordeaux}
         \and
              Institut f\"ur Astrophysik der Universit\"at Wien, 
              T\"urkenschanzstrasse 17, 1180 Wien, Austria
              \label{inst:Wien}
         \and
              Department of Theoretical Physics and Astrophysics, Masaryk University,
              Kotl\'a\v{r}sk\'a 267/2, 611\,37 Brno, Czech Republic
              \label{inst:Brno}
             }

   \date{Received 29 August 2013 / Accepted 20 October 2013}

 
  \abstract
  {Open clusters are an important tool for studying the chemical evolution of the Galactic disk. Metallicity estimates are available for about ten percent of the currently known open clusters. These metallicities are based on widely differing methods, however, which introduces unknown systematic effects.}
   {In a series of three papers, we investigate the current status of published metallicities for open clusters that were derived from a variety of photometric and spectroscopic methods. The current article focuses on spectroscopic methods. The aim is to compile a comprehensive set of clusters with the most reliable metallicities from high-resolution spectroscopic studies. This set of metallicities will be the basis for a calibration of metallicities from different methods.}
   {The literature was searched for [Fe/H] estimates of individual member stars of open clusters based on the analysis of high-resolution spectra. For comparison, we also compiled [Fe/H] estimates based on spectra with low and intermediate resolution.
At medium and high resolution, we found that differences in the analysis methods have a stronger effect on metallicity than quality differences in the observations.
   We retained only highly probable cluster members and introduced a restriction on atmospheric parameters.}
   {We combined 641 individual metallicity values for 458 stars in 78 open clusters from 86 publications to form our final set of high-quality cluster metallicities. The photometric metallicities discussed in the first paper of this series are systematically lower than the spectroscopic ones by about 0.1~dex, and the differences show a scatter of about 0.2~dex. In a preliminary comparison of our spectroscopic sample with models of Galactic chemical evolution, none of the models predicts the observed radial metallicity gradient.}
   {Photometric metallicities show a large intrinsic dispersion, while the more accurate spectroscopic sample presented in this paper comprises fewer than half the number of clusters. Only a sophisticated combination of all available photometric and spectroscopic data will allow us to trace the metallicity distribution in the Galactic disk on a local and global scale.}


   \maketitle
%

\section{Introduction}
\label{sect:introduction}


The metallicity of stars is the key to exploring of the chemical structure and evolution of the Galaxy. But which is the best way to determine this important astrophysical parameter?
High-resolution spectroscopy is often claimed to be the most accurate method \mbox{\citep[e.g.][]{2009A&A...494...95M}}. It is certainly true that this is the only method available to measure abundances of individual chemical elements. However, the overall stellar metallicity can be reliably inferred by a variety of other means provided that a good calibration is available. Furthermore, the accuracy of high-resolution spectroscopy is not always assessed in a satisfactory way.

Photometry and low-resolution spectroscopic studies have the advantage of providing results for large samples within a short time (both the observations and the analysis require less time). High-resolution spectra contain significantly more information and are therefore expected to give more accurate results. 
However, interpreting these data is based on theoretical stellar atmospheres and modelling of spectral lines. The complete procedure of an abundance analysis requires one to specify a considerable amount of input data, assumptions for the physics of relevant processes, and a significant number of free parameters. This leads to a wide variety of possibilities to analyse a given data set, which may lead to differences in the metallicity scales published by different research groups.

We set out to investigate to which extent this is the case, thus testing the \emph{accuracy} of spectroscopic stellar metallicities. We note that individual metallicity studies might well be able to achieve a high \emph{precision} for a limited number of stars \mbox{\citep[e.g.][]{2009ApJ...704L..66M}}, allowing one to draw important conclusions on a small subsystem of Galactic stars. But to obtain a complete picture of the Galactic chemical evolution requires one to combine the results of different authors, which will each be subject to systematic uncertainties of unknown magnitude.
For recent studies that compared spectroscopic analyses of well-known field stars using several different methods see for instance \mbox{\citet{lebzelter2012}} and Jofr\'e et al. (A\&A, submitted).

We focus here on metallicity determinations of stars in open clusters (OCs).
These objects have been used for a long time to investigate the radial metallicity gradient of the Galactic disk. Furthermore, the studied stars cover a wide range in stellar parameters (\Teff\ and \logg). Stars in any one cluster all formed at the same time from the same material with a unique metallicity, which provides an additional test for the consistency of the determined metallicities. 

Since the 1990s, we have seen a surge of spectroscopic metallicity studies of OC stars. This probably reflects the progress in automated abundance analysis, but is certainly also an effect of the availability of efficient multi-object spectrographs, such as FLAMES at ESO's VLT or Hydra at the WIYN telescope on Kitt Peak.
Open clusters are regarded as ideal probes for the chemical evolution of the Galactic disk, because they are numerous (more than 2000 presently known), are located throughout the disk, and span a wide range of ages \mbox{\citep{2002A&A...389..871D}}. 
An accurate knowledge of the cluster metallicity is of twofold importance. It obviously provides a measurement point for the disk chemistry at a certain location. Second, it is important for determining the cluster age and distance (using metallicity-dependent isochrones) and thus the point in time corresponding to the particular metallicity measurement.
When interpreting these measurements, it is important, however, to consider possible motions of the clusters during their lifetime \citep[e.g.][]{2009MNRAS.399.2146W}.

In spite of the astrophysical significance, metallicity estimations for OCs are still rare. In a previous paper, we compiled photometric metallicity determinations for 188 OCs \mbox{\citep[][hereafter Paper~I]{2010AaA...517A..32P}}.
Recent compilations of spectroscopic metallicities include those of \mbox{\citet[][63 clusters]{2009A&A...494...95M,2010AaA...523A..11M}} and \mbox{\citet[][89 clusters]{2011AaA...535A..30C}}.
These contain 29 clusters that are not included in our photometric sample, and thus the total current fraction of known clusters with a metallicity assessment is about 10\%. The two spectroscopic compilations take different approaches to arrive at a metallicity value for each cluster. Magrini et al. selected the result of one specific publication per cluster for their sample, while \mbox{\citet{2011AaA...535A..30C}} averaged all available determinations for each cluster. Both approaches suffer from the inhomogeneity inherent in data originating from many different research groups.

There are several recent attempts of constructing homogeneous sets of spectroscopic OC metallicities. 
\mbox{\citet{2010AJ....139.1942F}} complemented their own sample of eleven clusters with results from only three other groups and presented careful evaluations of possible differences in measurements and methods among all groups. This approach resulted in a homogeneous but small sample of 26 clusters.
For the BOCCE project\footnote{\url{http://www.bo.astro.it/~angela/bocce.html}} \mbox{\citep{2006AJ....131.1544B,2009IAUS..254..227B}} 
45 clusters have been selected, for which age, distance, reddening, and metallicity are being determined in a homogeneous way. For most of the clusters they obtained their own photometry, and for a large portion high-resolution spectra were acquired using only a few instruments. Metallicities determined in a consistent way are published for eight clusters so far. 

The most ambitious current effort is the \emph{Gaia-ESO} Public Spectroscopic Survey \citep[][PIs S. Randich and G. Gilmore]{2012Msngr.147...25G}, which will obtain medium- and high-resolution spectra for about $10^5$ Galactic stars during five years with one instrument (FLAMES-GIRAFFE-UVES). The target list includes stars in about 100 OCs, and major efforts are put into the preparation of a homogeneous analysis of this unprecedented dataset in a unique collaboration across more than ten research groups.

In the current paper, we aim to harvest the existing literature in the best possible way. The motivation is to obtain an up-to-date overview of the status and current limitations of OC metallicities. This is of crucial importance for the implementation of the on-going surveys (in particular the \emph{Gaia} space mission and the \emph{Gaia-ESO} survey) -- concerning both the best selection of target clusters and target stars, and the selection of clusters and stars for calibration purposes. Our approach is to compile atmospheric parameters and spectroscopic metallicities for individual stars in each cluster and compare the results of different authors for stars in common.

The article is arranged as follows: in Sect.~\ref{sect:data} we describe the selection of the metallicity determinations. In Sect.~\ref{sect:assessment} we compare the results obtained by different authors who studied the same OCs, star-by-star as well as for the mean cluster metallicity. We also assess the importance of spectrum quality for mean cluster metallicity.
In Sect.~\ref{sect:final} we present our final high-resolution sample. In Sect.~\ref{sect:discussion} we compare our sample with others and discuss possible applications of our sample to the study of Galactic structure, and Sect.~\ref{sect:conclusions} concludes the paper.


\section{Data selection}
\label{sect:data}


\subsection{High-resolution sample} 
\label{sect:hires}

To build a list of reference OCs with the most reliable metallicities, we first gathered individual stars -- highly probable members of OCs -- with atmospheric parameters determined from spectra of high resolution ($R=\lambda/\Delta\lambda$) and high signal-to-noise ratio (S/N). The lower limit in spectrum quality was set to $R$=25000, and S/N=50. We searched the PASTEL database \mbox{\citep{2010A&A...515A.111S}} and the recent literature for such stars in references posterior to 1990 and until June 2013. Only stars with \Teff $<$ 7000~K were included to avoid rapid rotators and chemical peculiarities.
All determinations not in PASTEL at the time of writing will be included in the database in the next update. We eliminated confirmed non-members, spectroscopic binaries, and chemically peculiar stars and kept only stars with a high probability of membership. Membership information was mainly based on radial-velocity criteria presented in \mbox{\citet{2008AaA...485..303M,2009AaA...498..949M}}. Criteria presented in the articles from which we gathered the spectroscopic determinations and information extracted from the WEBDA\footnote{\url{http://webda.physics.muni.cz}} \citep{2003A&A...410..511M} and Simbad databases were also used for membership evaluation. We started with a list of 571 stars in 86 OCs, with 830 metallicity determinations from 94 papers.
In Table~\ref{tab:appendix_members} we list the basic information for the full starting sample of cluster members, which should be sufficient to extract their parameters from the PASTEL catalogue (accessible via VizieR\footnote{\url{http://vizier.u-strasbg.fr/viz-bin/VizieR?-source=B\%2Fpastel}}).

\onecolumn
\begin{longtab}
\centering
\begin{longtable}{rlllcll}
\caption{\label{tab:appendix_members} List of the full starting sample of cluster members described in Sect.~\ref{sect:hires}.} \\
\hline\hline
\noalign{\smallskip}
S & Cluster ID & RA (J2000) & DEC (J2000) & Vmag & Bibcode & Star ID \\
\noalign{\smallskip}
\hline
\noalign{\smallskip}
\endfirsthead
\caption{ continued.} \\
\hline\hline
\noalign{\smallskip}
S & Cluster ID & RA (J2000) & DEC (J2000) & Vmag & Bibcode & Star ID \\
\noalign{\smallskip}
\hline
\noalign{\smallskip}
\endhead
\noalign{\smallskip}
\hline
\endfoot
\noalign{\smallskip}
\hline\hline
\noalign{\smallskip}
\multicolumn{7}{l}{\parbox{1.00\textwidth}{Column descriptions: The first column identifies the measurements that are included in the final high-resolution sample (S=2, Sect.~\ref{sect:final}; S=1 for the measurements not included in the final sample). The \emph{Bibcode} column gives the ADS Bibcode of the article from which the metallicity determination was taken. The last column gives a star identification, which is resolvable by Simbad, except for five stars in NGC~6791 measured by \citet{2012ApJ...756L..40G}, for which the numbers from \citet{2003PASP..115..413S} are given.}}\\
\endlastfoot
   1  & Berkeley17           & 05 20 23.875    & +30 37 21.91    & 14.49   & 2013AJ....145..107J & Cl* Berkeley 17 PJM 1035      \\
   1  & Berkeley17           & 05 20 23.875    & +30 37 21.91    & 14.49   & 2005AJ....129.2725F & Cl* Berkeley 17 PJM 1035      \\
   2  & Berkeley17           & 05 20 47.748    & +30 31 40.84    & 14.91   & 2005AJ....129.2725F & Cl* Berkeley 17 PJM 265       \\
   2  & Berkeley17           & 05 20 47.748    & +30 31 40.84    & 14.91   & 2013AJ....145..107J & Cl* Berkeley 17 PJM 265       \\
   1  & Berkeley17           & 05 20 36.500    & +30 30 35.18    & 14.09   & 2013AJ....145..107J & Cl* Berkeley 17 PJM 569       \\
   1  & Berkeley17           & 05 20 36.500    & +30 30 35.18    & 14.09   & 2005AJ....129.2725F & Cl* Berkeley 17 PJM 569       \\
   2  & Berkeley18           & 05 22 13.8      & +45 27 58.9     & 15.79   & 2012AJ....144...95Y & Cl* Berkeley 18 K 1163        \\
   2  & Berkeley18           & 05 22 13.8      & +45 27 58.9     & 15.79   & 2013AJ....145..107J & Cl* Berkeley 18 K 1163        \\
\end{longtable}
\end{longtab}
%
\twocolumn

For comparison purposes, we also extracted mean cluster metallicities from a number of studies (post-1990) at high resolution and low S/N ($<50$), and also at medium resolution ($R<25000$, high and low S/N), that is, below the quality criteria defined above. Some of these determinations are discussed in Sect.~\ref{sect:metalrich} at the individual star level, and several are included in Sect.~\ref{sect:mean} at the cluster level. The remaining determinations are not further discussed in this paper. For future reference, we list them in Table~\ref{tab:appendix_lowerquality}.

\begin{table*}
\caption{Mean cluster metallicities and references for lower-quality studies not discussed in the paper (not in Tables~\ref{tab:mean1} or \ref{tab:mean2}).}
\label{tab:appendix_lowerquality}
\centering
\begin{tabular}{lrrrrll}
\hline\hline
\noalign{\smallskip}
\multicolumn{7}{l}{High-resolution, low SNR ($<50$) studies} \\
\noalign{\smallskip}
\hline\hline
\noalign{\smallskip}
Cluster ID & mean \FeH & std. dev. & resolution & \# & star type & reference \\
\noalign{\smallskip}
\hline
\noalign{\smallskip}
Berkeley 21  &   $-$0.54 & 0.20  &  48000 &  3   &    giant  & \citet{1999AaA...348L..21H} \\
Berkeley 22  &   $-$0.32 & 0.19  &  34000 &  2   &    giant  & \citet{2005AJ....130..652V} \\
Berkeley 25  &   $-$0.20 & 0.05  &  40000 &  4   &    giant  & \citet{2007AaA...476..217C} \\
Berkeley 66  &   $-$0.48 & 0.24  &  34000 &  2   &    giant  & \citet{2005AJ....130..652V} \\
Berkeley 73  &   $-$0.22 & 0.10  &  40000 &  2   &    giant  & \citet{2007AaA...476..217C} \\
Berkeley 75  &   $-$0.22 & 0.20  &  40000 &  1   &    giant  & \citet{2007AaA...476..217C} \\
NGC 2355     &   $-$0.07 & 0.11  &  42000 &  15  &    giant  & \citet{2000AaA...357..484S} \\
Ruprecht 4   &   $-$0.09 & 0.05  &  40000 &  3   &    giant  & \citet{2007AaA...476..217C} \\
Ruprecht 7   &   $-$0.26 & 0.05  &  40000 &  5   &    giant  & \citet{2007AaA...476..217C} \\
\noalign{\smallskip}
\hline\hline
\noalign{\smallskip}
\multicolumn{7}{l}{Medium-resolution, high SNR ($>50$) studies} \\
\noalign{\smallskip}
\hline\hline
\noalign{\smallskip}
Cluster ID & mean \FeH & std. dev. & resolution & \# & star type & reference \\
\noalign{\smallskip}
\hline
IC 2581     &    $-$0.34 &       &  18000        &   1   &    giant        &   \citet{1994ApJS...91..309L} \\
IC 4725     &       0.17 & 0.09  &  18000        &   3   &    giant        &   \citet{1994ApJS...91..309L} \\
NGC 2168    &    $-$0.21 & 0.10  &  20000        &   9   &    dwarf        &   \citet{2001ApJ...549..452B} \\
NGC 2425    &    $-$0.15 & 0.09  &  21000        &   4   &    giant        &   \citet{2011AJ....142...59J} \\
NGC 3293    &       0.14 & 0.11  &  $\sim$20000  &   2   &    giant+dwarf  &   \citet{2007AaA...471..625T} \\
NGC 4755    &       0.39 & 0.25  &  $\sim$20000  &   1   &    dwarf        &   \citet{2007AaA...471..625T} \\
NGC 6067    &       0.01 & 0.12  &  18000        &   7   &    giant        &   \citet{1994ApJS...91..309L} \\
NGC 6087    &    $-$0.01 & 0.23  &  18000        &   3   &    giant        &   \citet{1994ApJS...91..309L} \\
NGC 6611    &       0.17 & 0.15  &  20000        &   1   &    dwarf        &   \citet{2007AaA...471..625T} \\
NGC 6882/5  &    $-$0.02 & 0.01  &  18000        &   2   &    giant        &   \citet{1994ApJS...91..309L} \\
Tombaugh 2  &    $-$0.28 & 0.08  &  21000        &   7   &    giant        &   \citet{2008MNRAS.391...39F} \\
Tombaugh 2  &    $-$0.31 & 0.02  &  $\sim$17000  &  13   &    giant        &   \citet{2010AaA...509A.102V} \\
\noalign{\smallskip}
\hline\hline
\noalign{\smallskip}
\multicolumn{7}{l}{Medium-resolution, low SNR ($<50$) studies} \\
\noalign{\smallskip}
\hline\hline
\noalign{\smallskip}
Cluster ID & mean \FeH & std. dev. & resolution & \# & star type & reference \\
\noalign{\smallskip}
\hline
NGC 1883    &    $-$0.20 & 0.22  &  20000 &  2   &    giant &  \citet{2007MNRAS.379.1089V} \\
NGC 6253    &       0.36 & 0.20  &  15000 &  2   &    giant &  \citet{2000ASPC..198..273C} \\
NGC 6791    &       0.40 & 0.10  &  20000 &  1   &    BHB   &  \citet{1998ApJ...502L..39P} \\
NGC 6791    &       0.39 & 0.05  &  20000 &  10  &    giant &  \citet{2006ApJ...643.1151C} \\
\noalign{\smallskip}
\hline\hline
\end{tabular}
\end{table*}
%

\begin{table*}
\caption{Mean cluster metallicities for low-resolution studies of giant stars in clusters not discussed in the paper (not in Table~\ref{tab:mean1}). The references are \citet{2002AJ....124.2693F,2003PASP..115...96W,2005AJ....130.1916M}.}
\label{tab:appendix_lowres}
\centering
\begin{tabular}{lrrrllrrrl}
\noalign{\smallskip}
\hline\hline
\noalign{\smallskip}
Cluster ID & mean \FeH & std. dev. & \# & first author & Cluster ID & mean \FeH & std. dev. & \# & first author\\
\noalign{\smallskip}
\hline
Berkeley 17   &  -0.33 &  0.12 & 13    &  Friel    & NGC 2099      &  0.05  &  0.14 & 8     &  Marshall \\
Berkeley 20   &  -0.61 &  0.14 & 6     &  Friel    & NGC 2141      &  -0.33 &  0.10 & 6     &  Friel    \\
Berkeley 21   &  -0.62 &  0.11 & 4     &  Friel    & NGC 2324      &  -0.06 &  0.07 & 4     &  Marshall \\
Berkeley 31   &  -0.40 &  0.16 & 17    &  Friel    & NGC 2324      &  -0.15 &  0.16 & 7     &  Friel    \\
Berkeley 32   &  -0.50 &  0.04 & 10    &  Friel    & NGC 2360      &  -0.26 &  0.02 & 4     &  Friel    \\
Collinder 261 &  -0.16 &  0.13 & 21    &  Friel    & NGC 2477      &  -0.13 &  0.10 & 28    &  Friel    \\
IC 166        &  -0.34 &  0.16 & 4     &  Friel    & NGC 2506      &  -0.44 &  0.06 & 5     &  Friel    \\
King 5        &  -0.30 &  0.17 & 19    &  Friel    & NGC 2539      &  -0.04 &  0.05 & 4     &  Marshall \\
King 8        &  -0.39 &       & 1     &  Friel    & NGC 3680      &  -0.19 &  0.05 & 7     &  Friel    \\
King 11       &  -0.27 &  0.15 & 16    &  Friel    & NGC 3960      &  -0.34 &  0.09 & 5     &  Friel    \\
Melotte 66    &  -0.47 &  0.09 & 4     &  Friel    & NGC 6819      &  -0.11 &  0.06 & 7     &  Friel    \\
NGC 188       &  -0.10 &  0.09 & 21    &  Friel    & NGC 6819      &  0.07  &  0.24 & 4     &  Marshall \\
NGC 188       &  0.08  &  0.05 & 14    &  Worthey  & NGC 6940      &  -0.12 &  0.10 & 6     &  Friel    \\
NGC 752       &  -0.18 &  0.04 & 9     &  Friel    & Pismis 2      &  -0.07 &  0.23 & 9     &  Friel    \\
NGC 1193      &  -0.51 &  0.09 & 4     &  Friel    & Tombaugh 2    &  -0.44 &  0.09 & 12    &  Friel    \\
\noalign{\smallskip}
\hline\hline
\end{tabular}
\end{table*}

\subsection{Low-resolution sample} 
As low-resolution spectroscopic metallicity investigations we considered spectra taken at $R \approx 1000\,-\,2000$, from which spectroscopic indices, that is, the strength of absorption features, were measured (predominantly Fe~I and Fe peak blends). A calibration giving the index strength as a function of the atmospheric parameters allows one to determine the metallicity. Low-resolution studies of OC stars are relatively rare in the literature. In addition to a query within ADS, the bibliography of the OC database WEBDA was used, and references in high-resolution studies were checked in this respect.
We found \mbox{\citet{1993AaA...267...75F}}, \mbox{\citet{1993PASP..105.1253T}}, \mbox{\citet{2002AJ....124.2693F}}, \mbox{\citet{2003PASP..115...96W}}, and \mbox{\citet{2005AJ....130.1916M}}, the first three of which claim to be on the same metallicity scale. Since \mbox{\citet{2002AJ....124.2693F}} revised the results of the former two studies in combination with new data, we included only their metallicity determinations of 39 OCs in total. Additionally, we adopted the result for the cluster NGC~6705, investigated by \mbox{\citet{1993PASP..105.1253T}}, which is not listed in \mbox{\citet{2002AJ....124.2693F}} due to their restriction to ages older than 0.7~Gyr.
\mbox{\citet{2002AJ....124.2693F}} and \mbox{\citet{2005AJ....130.1916M}} determined \FeH\ based on spectroscopic indices defined in \mbox{\citet{1987AJ.....93.1388F}}\footnote{Six and eleven indices were used in the two works, respectively.}. \mbox{\citet{2003PASP..115...96W}} used seven indices on the Lick/IDS system defined in \mbox{\citet{1994ApJS...94..687W}}.

\mbox{\citet{2003PASP..115...96W}}, and \mbox{\citet{2005AJ....130.1916M}} seem to present a higher metallicity scale than \mbox{\citet{2002AJ....124.2693F}}. \mbox{\citet{2003PASP..115...96W}} studied two clusters (NGC~188 and NGC~6791) and obtained a metallicity about 0.2~dex higher than \mbox{\citet{2002AJ....124.2693F}} for both clusters. \mbox{\citet{2005AJ....130.1916M}} have three clusters (out of seven) in common with \mbox{\citet{2002AJ....124.2693F}}, indicating a similar tendency (an offset between 0.1 and 0.2~dex). The only exception is NGC~6705, for which the result of \mbox{\citet{2005AJ....130.1916M}} is slightly lower than that of \mbox{\citet[][they agree within the errors]{1993PASP..105.1253T}}.

In total, we found 49 metallicity values for 43 individual clusters investigated with low-resolution spectroscopy, of which 34 are also included in the high-resolution sample.
Four of the low-resolution clusters have neither photometric nor high-resolution spectroscopic metallicities.
Some of these determinations are discussed in Sects.~\ref{sect:m67} to \ref{sect:metalpoor} and in Sect.~\ref{sect:mean}. The remaining determinations are not discussed in this paper. For future reference, we list them in Table~\ref{tab:appendix_lowres}.
A detailed comparison of these results, and if possible, a recalibration by means of the high-resolution sample, is planned for the next paper in this series. 



\section{Assessment of spectroscopic metallicities}
\label{sect:assessment}

\begin{figure*}[t]
\includegraphics[width=\textwidth,trim=0 180 0 280,clip]{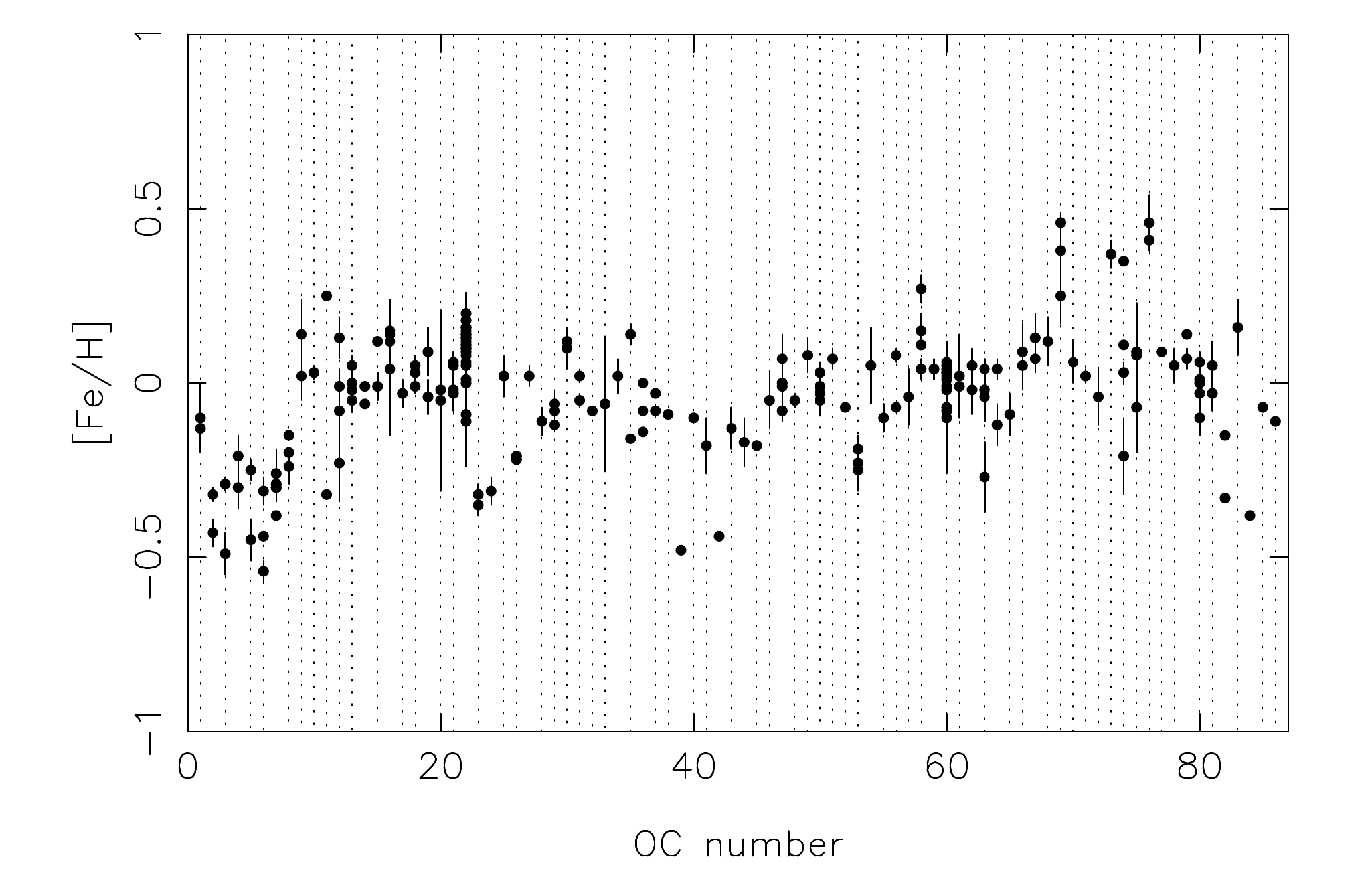}
\caption{Mean metallicity of OCs per publication, as listed in Table \ref{t:mean_paper}.}
\label{f:mean_paper}
\end{figure*}
%


\subsection{Mean metallicities and quoted errors for the high-resolution sample}
\label{sect:errors}

A significant number of individual OCs in the high-resolution sample were studied by different authors with different instruments, methods, and line lists.
Throughout this article, we use the term \emph{metallicity} synonymously with \emph{iron abundance}, [Fe/H]. Different authors used different approaches for determining the iron abundances -- using either only Fe~I lines or only Fe~II lines, or both. If abundances for both ions are given, we used the metallicity value as given by the authors, which can be either an unweighted mean of Fe~I and Fe~II abundances or a mean weighted by the number of lines used in each case. In the few cases where metallicity values for individual stars are not clearly stated, we formed an unweighted mean of Fe~I and Fe~II abundances. 
In the majority of the publications the metallicity is based on Fe~I only. In about 10\% of the cases a weighted mean is given. Another 10\% achieved ionization equilibrium in the analysis by adapting the atmospheric parameters to such an extent that the Fe~I and Fe~II abundances agree exactly.
There is also a strong variation in the number of Fe lines used for the analysis, which can be anything between two and 130 Fe~I lines (and individual publications with 180 or 265 lines), and one to 15 Fe~II lines (40 lines in one publication).

These variations are partly due to the different telescopes and instruments used to obtain the observed spectra, in particular their wavelength coverage.
The observations were obtained at about 30 different telescopes, most of them in the 2--4~m class, and a few at 8--10~m.
Echelle spectrographs were used for most observations, providing a large wavelength coverage, but settings focussing on different parts of the optical region were used. The minimum wavelength of the spectra varies between publications from 360 to 700~nm, and the maximum wavelength from 600 to 1060~nm. 

\begin{figure}
\includegraphics[width=\columnwidth,trim=40 40 60 60,clip]{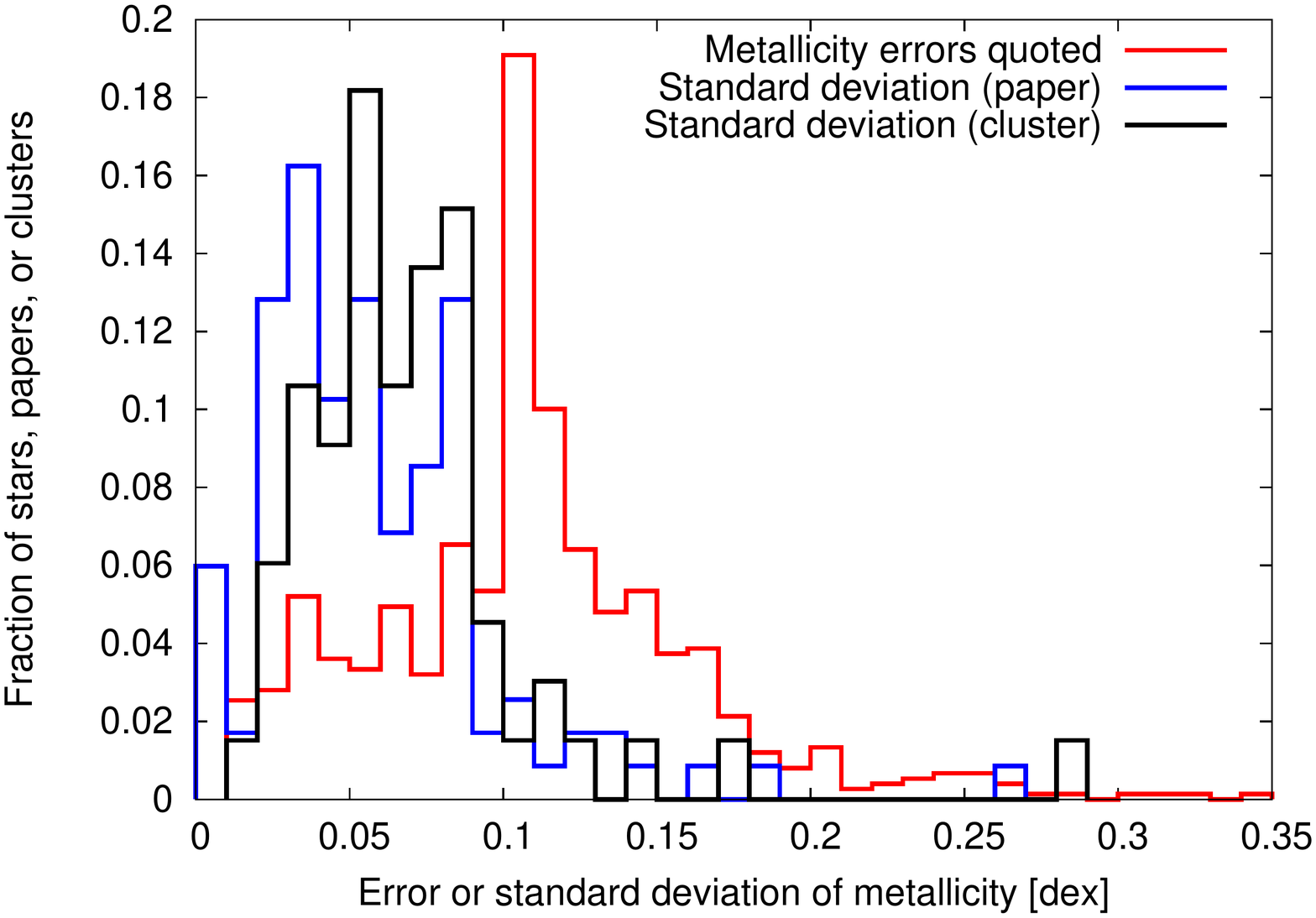}
\caption{Distribution of metallicity errors quoted by authors (red line) compared with the distributions of standard deviations for mean metallicities per cluster for each paper (blue line), and for all papers (black line), combining determinations for at least three stars.}
\label{f:errorhistogram}
\end{figure}

We show the  individual metallicities averaged by OC and by reference in Fig.~\ref{f:mean_paper} and list them in Table~\ref{t:mean_paper}.
The figure shows a certain lack of homogeneity for the clusters studied by several authors.
The errors on individual [Fe/H] determination quoted by the authors are typically around 0.1~dex, and for most determinations they are less 0.2~dex (see Fig.~\ref{f:errorhistogram}).
Only 14 determinations for seven clusters in six papers have quoted metallicity errors between 0.25 and 0.35~dex
(Berkeley~22, \mbox{\citealt{2013AJ....145..107J}};
Berkeley~29 and NGC~2141, \mbox{\citealt{2005AJ....130..597Y}}, see also Sect.~\ref{sect:other};
Collinder~261, \mbox{\citealt{2003AJ....126.2372F}}, see also Sect.~\ref{sect:peculiar};
NGC~2112, \mbox{\citealt{1996AJ....112.1551B}}, see also Sect.~\ref{sect:mean};
NGC~3680, \mbox{\citealt{2001AaA...374.1017P}}, see also Sect.~\ref{sect:metalpoor};
NGC~6705, \mbox{\citealt{2012AaA...538A.151S}}).
For 81 determinations, the authors do not quote any error for the metallicity.
There is a weak dependence of errors on \Teff\ and \FeH, such that the largest errors are quoted for stars with \Teff$<$5000~K and \FeH$<-0.1$~dex.

We note that the errors quoted by most authors are in fact the standard deviations of the abundances determined from the selected Fe lines, with the exception of \mbox{\citet{2010AaA...511A..56P}} and \mbox{\citet{2011AaA...535A..30C}}. These authors quote the standard error of the line abundances, and we multiplied their errors by the square root of the number of lines used. The uncertainty in stellar parameters causes an additional uncertainty in the Fe abundances. This type of external error is typically estimated to be 0.1~dex in the publications included in this work.
An additional source of systematic differences between different studies might arise from the choice of the solar reference metallicity. We did not assess the extension of this effect for each individual paper. However, for the studies that are based on a differential analysis with respect to a solar spectrum or that use astrophysical oscillator strengths, the derived or adopted value of the solar Fe abundance is not important. We expect this to be the case for the majority of the publications. The remaining studies may be affected by external errors of up to 0.1~dex.

Standard deviations of the mean cluster metallicity from each individual paper are mostly lower than 0.1~dex, with a peak below 0.05~dex, which shows that the internal uncertainties of metallicity determinations are probably lower than the quoted errors (see Fig.~\ref{f:errorhistogram}). When we combine all determinations per cluster from all papers, the peak of the standard deviations is slightly higher than 0.05~dex, which suggests that external errors are inherent in the datasets.
In the following sections, we investigate some cases with a large number of metallicity determinations by different authors, including low-resolution studies, and some cases with large standard deviations around the mean metallicity. 

In the high-resolution sample, there are 26 OCs for which metallicity determinations are available for fewer than three stars. The reliability of the metallicities for these clusters is difficult to assess. We will apply the conclusions drawn from the comparisons for the well-studied clusters\footnote{The starting sample contains 47 clusters with three to ten stars, ten clusters with twelve to 20 stars, two clusters with 29 stars, and one with 76.} to these poor-studied ones for the selection of the final spectroscopic OC metallicity sample.

\onecolumn
\begin{longtab}
\begin{longtable}{  r   l   l   l   l   r   l  }
   \caption{\label{t:mean_paper} Weighted mean metallicity (column 5) of OCs, computed for each reference (column 7) giving atmospheric parameters based on high-resolution, high-S/N spectra. Only highly probable members and non-SBs were considered. See also Fig.~\ref{f:mean_paper}.} \\
\hline\hline
\noalign{\smallskip}
(1)    & (2)  & (3)          & (4)          & (5)        &   (6)& (7)       \cr
OC no. & Name & \Teff\ range & \logg\ range & Mean \FeH  &   \# & Reference \cr
\noalign{\smallskip}
\hline
\noalign{\smallskip}
\endfirsthead
\caption{ continued.} \\
\hline\hline
\noalign{\smallskip}
OC no. & Name & \Teff\ range & \logg\ range & Mean \FeH  &   \# & Reference \cr
\noalign{\smallskip}
\hline
\noalign{\smallskip}
\endhead
\noalign{\smallskip}
\hline
\endfoot
\noalign{\smallskip}
\hline\hline
\noalign{\smallskip}
\multicolumn{7}{l}{\parbox{1.00\textwidth}{The \Teff\ and \logg\ ranges and the number of determinations (\#) are given in columns 3, 4, and 6. In most cases, the latter corresponds to the number of stars analysed. In the case of \citet{2009AaA...493..309S} giant stars are counted twice because two different line lists were used. In the cases of \citet{2000AaA...360..499T} and \citet{2004AaA...424..951P}, for three of seven and three of 21 stars, respectively, two or three different spectra were used. The resulting values were treated as independent determinations.}}\\
\endlastfoot
 1 & Berkeley17         &    4300  4490 &  1.50  2.00 & $-$0.10 $\pm$ 0.10 &  3 &  \citet{2005AJ....129.2725F} \cr
 1 & Berkeley17         &    4300  4500 &  1.50  2.10 & $-$0.13 $\pm$ 0.03 &  3 &  \citet{2013AJ....145..107J} \cr
 2 & Berkeley18         &    4400  4650 &  1.90  2.30 & $-$0.32 $\pm$ 0.02 &  2 &  \citet{2013AJ....145..107J} \cr
 2 & Berkeley18         &    4400  4500 &  1.90  2.20 & $-$0.43 $\pm$ 0.04 &  2 &  \citet{2012AJ....144...95Y} \cr
 3 & Berkeley20         &    4400  4700 &  1.70  2.10 & $-$0.29 $\pm$ 0.02 &  2 &  \citet{2008AaA...488..943S} \cr
 3 & Berkeley20         &    4500  4590 &  1.80  2.20 & $-$0.49 $\pm$ 0.06 &  2 &  \citet{2005AJ....130..597Y} \cr
 4 & Berkeley21         &    4520  4650 &  1.90  2.10 & $-$0.21 $\pm$ 0.06 &  2 &  \citet{2013AJ....145..107J} \cr
 4 & Berkeley21         &    4500  4625 &  1.70  1.90 & $-$0.30 $\pm$ 0.06 &  2 &  \citet{2012AJ....144...95Y} \cr
\end{longtable}
\end{longtab}
\twocolumn

\subsection{Solar-metallicity cluster: M67}
\label{sect:m67}

   \begin{figure*}
   \centering
   \includegraphics[width=\textwidth,trim=10 50 10 50,clip]{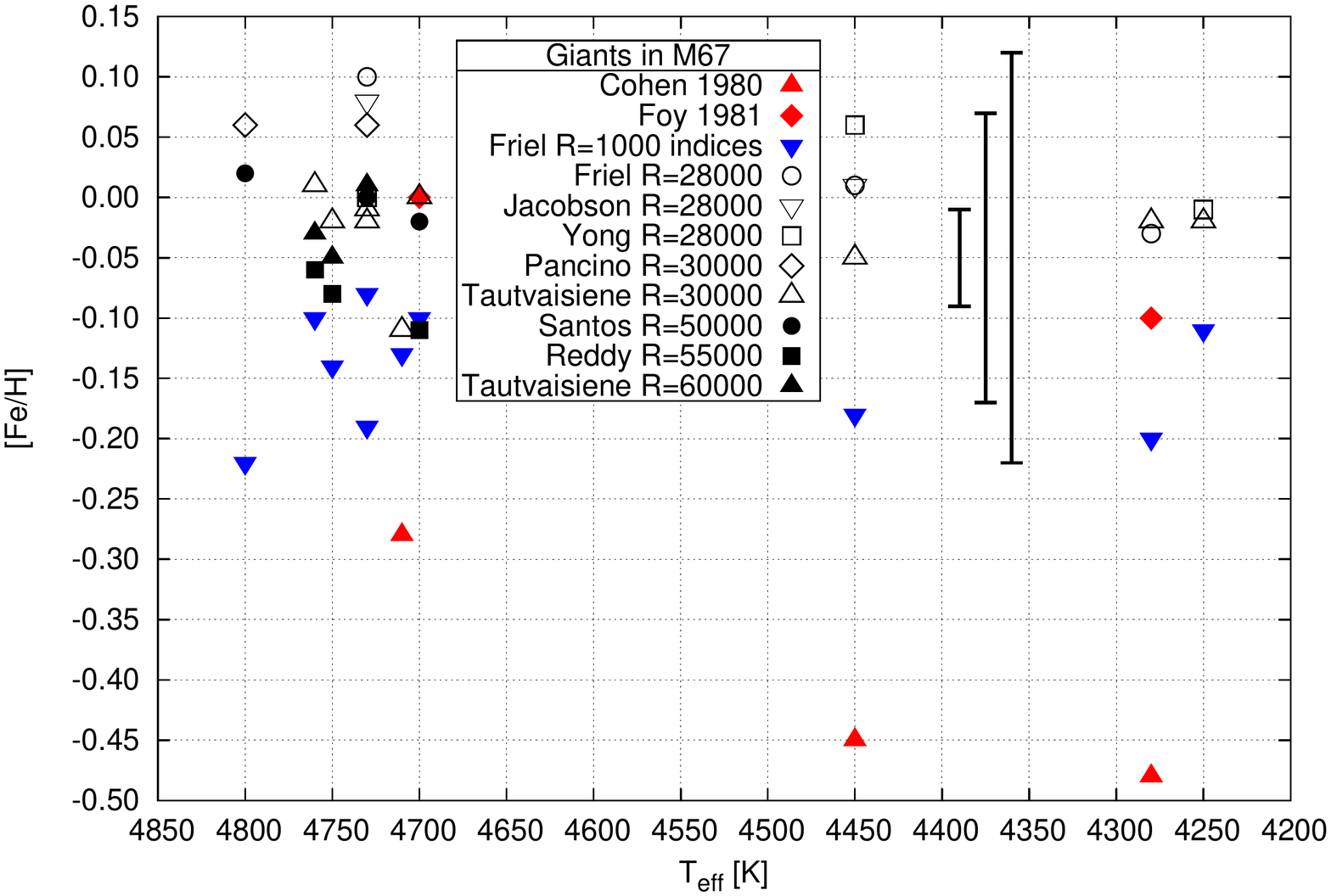}
   \caption{
   Metallicity versus reference \Teff\ for individual \emph{giant} stars in NGC~2682 (M67) that were analysed by more than one author using high- and low-resolution spectra.
The metallicities are taken from the following publications:
\citet{1980ApJ...241..981C,1981AaA....99..221F,2000AaA...360..499T,2002AJ....124.2693F,2005AJ....130..597Y,2009AaA...493..309S,2010AaA...511A..56P,2010AJ....139.1942F,2013AJ....145..107J}; and \citet{2013MNRAS.431.3338R}.
Data for the following stars are shown (reference \Teff, number from \citealt{1906PhDT.........1F}):
(4250, 108); (4280, 170); (4450, 105); (4700, 164); (4710, 224); (4730, 141); (4730, 266); (4750,  84); (4760, 151); and (4800, 286).
The bars at the upper right indicate minimum, mean, and maximum uncertainties  for the individual [Fe/H] values quoted in the publications.
   }
   \label{fig:M67giants}
   \end{figure*}

Metallicities for \emph{giant} stars (\logg $\le$ 3.0) in M67 (NGC~2682) have been determined in a considerable number of publications since 1990, including the low-resolution studies by \mbox{\citet{2002AJ....124.2693F}} and \mbox{\citet{2005AJ....130.1916M}}.
Seven studies published after 1990 are based on high-resolution, high-S/N spectra.
We plot the metallicities determined for ten individual giant stars by more than one author at high and low resolution in Fig.~\ref{fig:M67giants} as a function of reference \Teff\ (see figure caption for references).
The reference \Teff\ values were taken from \mbox{\citet{2000AaA...360..499T}} except for Fagerholm 286 (average of \mbox{\citealt{2009AaA...493..309S}} and \mbox{\citealt{2010AaA...511A..56P}}).
Each \Teff\ value corresponds to one star, except for \Teff=4730~K, which corresponds to two stars.
The reference \Teff\ values agree to within 100~K with those given in the individual publications.
The internal uncertainties for the individual \FeH\ values quoted by the authors range from 0.04 \mbox{\citep{2013MNRAS.431.3338R}} to 0.17~dex \mbox{\citep{2013AJ....145..107J}}, with a mean of 0.12~dex.
In addition, we show metallicity determinations for four of these stars from two studies made before CCD detectors became available \mbox{\citep{1980ApJ...241..981C,1981AaA....99..221F}}, which are based on somewhat lower-resolution spectra, and atmospheric models and atomic data available at that time.

It is obvious that the metallicity determined from spectroscopic indices \mbox{\citep{2002AJ....124.2693F}} is on a more metal-poor scale than the high-resolution metallicities ($\sim$0.1~dex difference).
One of the older studies \mbox{\citep[][$R\approx$17000]{1980ApJ...241..981C}} shows a large systematic offset from the high-resolution studies, while the other one \mbox{\citep{1981AaA....99..221F}} agrees very well. 
All but five of the high-resolution metallicities are confined within $\pm 0.06$~dex, with no obvious dependence on resolution or \Teff.
\mbox{\citet[][Sect. 6.1.1]{2010AJ....139.1942F}} compared their results for three stars (Fagerholm 105, 141, and 170) with three other studies (for Fe as well as other elements), and arrived at similar conclusions.
The extended comparison presented here indicates that older abundance-analysis techniques applied to solar metallicity giant stars have larger systematic uncertainties than modern ones (post-1990). 

   \begin{figure*}
   \centering
   \includegraphics[width=\textwidth,trim=10 50 10 50,clip]{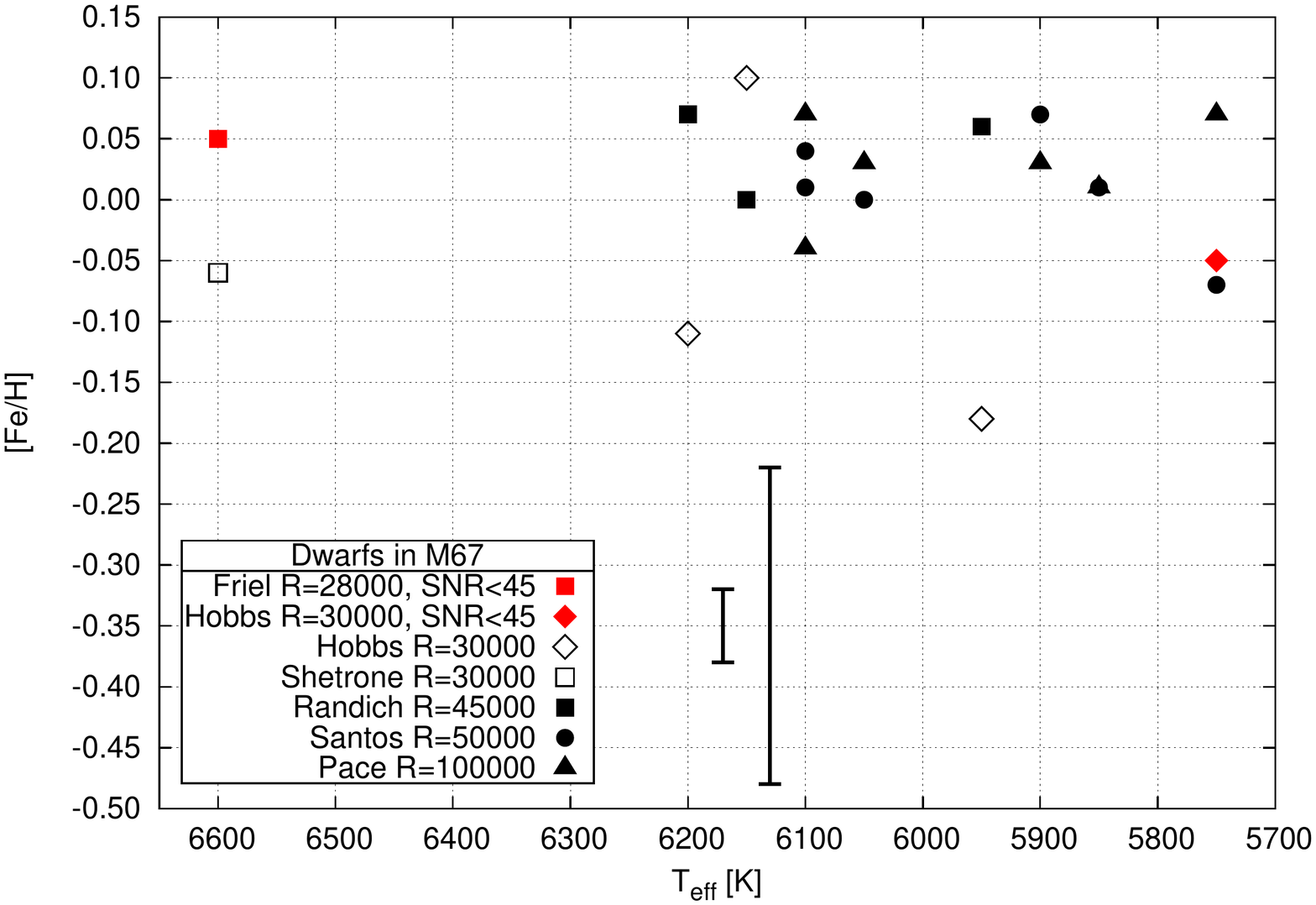}
   \caption{
   Same as Fig.~\ref{fig:M67giants}, for \emph{dwarf} stars in M67 (NGC~2682).
The metallicities are taken from the following publications:
\citet{1991AJ....102.1070H,1992ApJ...387..170F,2000AJ....120.1913S,2006AaA...450..557R,2008AaA...489..403P}; and \citet{2009AaA...493..309S}.
Data for the following stars are shown (reference \Teff, number from \citealt{1977A&AS...27...89S}):
(5750, 746); (5850, 1255); (5900, 1048); (5950, 1256); (6050, 1283); (6100, 1092); (6100, 1287); (6150, 994); (6200, 998); and (6600, 997).
The S/N of the spectra used in the different publications is higher than 50, except when stated otherwise.
The bars to the right of the legend indicate minimum and maximum uncertainties for the individual [Fe/H] values quoted in the publications.
   }
   \label{fig:M67dwarfs}
   \end{figure*}

For \emph{dwarf} stars in M67 we found metallicities in six publications, all based on high-resolution and high S/N spectra.
Ten dwarf stars were analysed by at least two authors. 
We plot the metallicities determined for these stars in Fig.~\ref{fig:M67dwarfs} as a function of reference \Teff\ on the same vertical scale as in Fig.~\ref{fig:M67giants} for the giant stars (see figure caption for references).
The reference \Teff\ values are averages of the determinations from different works.
Each \Teff\ value corresponds to one star, except for \Teff=6100~K, which corresponds to two stars (see figure caption).
The reference \Teff\ values agree to within 100~K with those given in the individual publications. The quoted uncertainties for the individual \FeH\ values range from 0.03 \mbox{\citep{2006AaA...450..557R}} to 0.13~dex \mbox{\citep{1991AJ....102.1070H}}.

Among the dwarf analyses, the oldest study by \mbox{\citet{1991AJ....102.1070H}} stands out among the others because it shows the largest dispersion.
Otherwise, the spread in abundances is the same as for the giant stars.
The lower-resolution metallicities agree with the others, but the small number of these points does not allow a general conclusion.
The star with the highest temperature (Sanders 997) is an M67 blue straggler and a probable spectroscopic binary star \mbox{\citep{2000AJ....120.1913S}}.
A comparison of dwarf and giant metallicities for this extraction from the starting sample of cluster members results in equal mean values within the standard deviations
(cf. Sect.~\ref{sect:final} for a comparison based on the full final sample). The mean of the 28 post-1990 high-resolution values for giants (full and open black symbols in Fig.~\ref{fig:M67giants}) is $-0.01\pm0.05$~dex, 
while the mean of the 15 most recent high-resolution values for dwarfs (full black symbols in Fig.~\ref{fig:M67dwarfs}) is 0.02$\pm$0.04~dex. 

\subsection{Metal-rich cluster: NGC~6791}
\label{sect:metalrich}

   \begin{figure*}
   \centering
   \includegraphics[width=\textwidth,trim=10 50 10 50,clip]{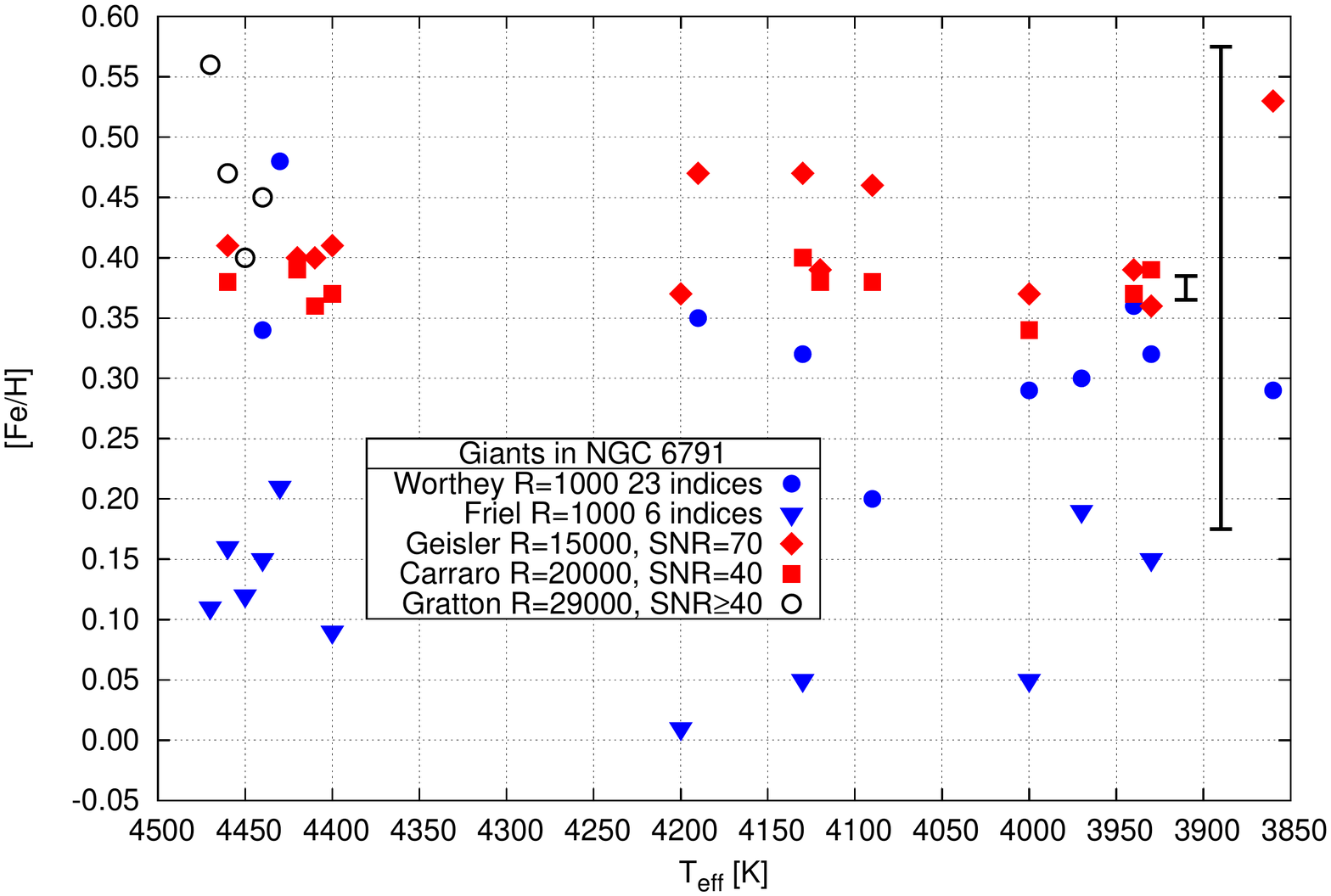}
   \caption{Same as Fig.~\ref{fig:M67giants}, for giant stars in NGC~6791.
The metallicities are taken from the following publications:
\citet{2003PASP..115...96W,2002AJ....124.2693F,2012ApJ...756L..40G,2006ApJ...643.1151C}; and \citet{2006ApJ...642..462G}.
Data for the following stars are shown (reference \Teff\ in K, star number from \citealt{2003PASP..115..413S}):
(3860,  8904);
(3930,  7972);
(3940,  8266);
(3970,  5342);
(4000, 11814);
(4090,  8563);
(4120, 10898);
(4130,  8988);
(4190,  4952);
(4200,  6288);
(4400,  3369);
(4410,  4715);
(4420,  7922);
(4430,  2723);
(4440,  9462);
(4450, 10806);
(4460,  8082);
and (4470,  9316).
The bars to the right indicate minimum and maximum uncertainties for the individual [Fe/H] values quoted in the publications.
   }
   \label{fig:NGC6791}
   \end{figure*}

For NGC~6791, we found five publications since 1990 with metallicity determinations for 18 giant stars appearing in at least two of them.
We plot the metallicities determined for these stars in Fig.~\ref{fig:NGC6791} as a function of reference \Teff\ on the same vertical and horizontal scale as in Fig.~\ref{fig:M67giants} for the M67 giant stars (see figure caption for references).
The reference \Teff\ values are averages of the determinations from different works.
Each \Teff\ value corresponds to one star (see figure caption).
The reference \Teff\ values agree to within 100~K with those given in the individual publications.
The minimum and maximum uncertainties for the individual \FeH\ values were found in two low-resolution studies and represent the standard deviations of metallicities determined from several indices (0.01~dex in \mbox{\citealt{2003PASP..115...96W}}, and in 0.2~dex \mbox{\citealt{2002AJ....124.2693F}}).
There are no multiple metallicity determinations available for dwarf stars in this cluster.

Again, the values from spectroscopic indices by \mbox{\citet{2002AJ....124.2693F}} are lower than the others, by a larger amount than at solar metallicity ($\sim$0.3~dex).
However, the indices-based metallicities by \mbox{\citet{2003PASP..115...96W}} are close to the values derived from higher-resolution spectra.
There is one star in common between the medium- and high-resolution studies of \mbox{\citet{2012ApJ...756L..40G}}, \mbox{\citet{2006ApJ...643.1151C}}, and \mbox{\citet{2006ApJ...642..462G}}, where the highest-resolution study gives the highest metallicity. The error quoted by \mbox{\citet{2006ApJ...642..462G}} is twice as large as the difference, however, and thus the difference is not significant.
Nine additional stars are in common between \mbox{\citet{2012ApJ...756L..40G}} and \mbox{\citet{2006ApJ...643.1151C}}, and for all except two stars, the metallicities agree very well.
The mean metallicity values derived from the whole samples of these publications show differences similar to the standard deviations. \mbox{\citet{2012ApJ...756L..40G}} determined \FeH=0.42$\pm$0.05 for 16 stars, \mbox{\citet{2006ApJ...643.1151C}} determined \FeH=0.38$\pm$0.02 for ten stars, and \mbox{\citet{2006ApJ...642..462G}} determined \FeH=0.47$\pm$0.07 for four stars\footnote{Note that Stetson 8082 is not included in Table~\ref{t:mean_paper} because its S/N is lower than 50.}.
These differences are probably not caused by the different resolutions, because the analysis of five subgiants at $R=45000$ by \mbox{\citet{2012ApJ...756L..40G}} results in a mean metallicity that agrees with all three of the medium- and high-resolution studies of giants within the standard deviations (see Table~\ref{t:mean_paper}).
Reasons for the differences could be the different instruments used (Hydra at WIYN and SARG at TNG), different stellar samples, different spectrum synthesis codes, and different line lists (\mbox{\citealt{2006ApJ...642..462G}} used the broadest wavelength range).

\subsection{Metal-poor clusters}
\label{sect:metalpoor}

   \begin{figure*}
   \centering
   \includegraphics[width=\textwidth]{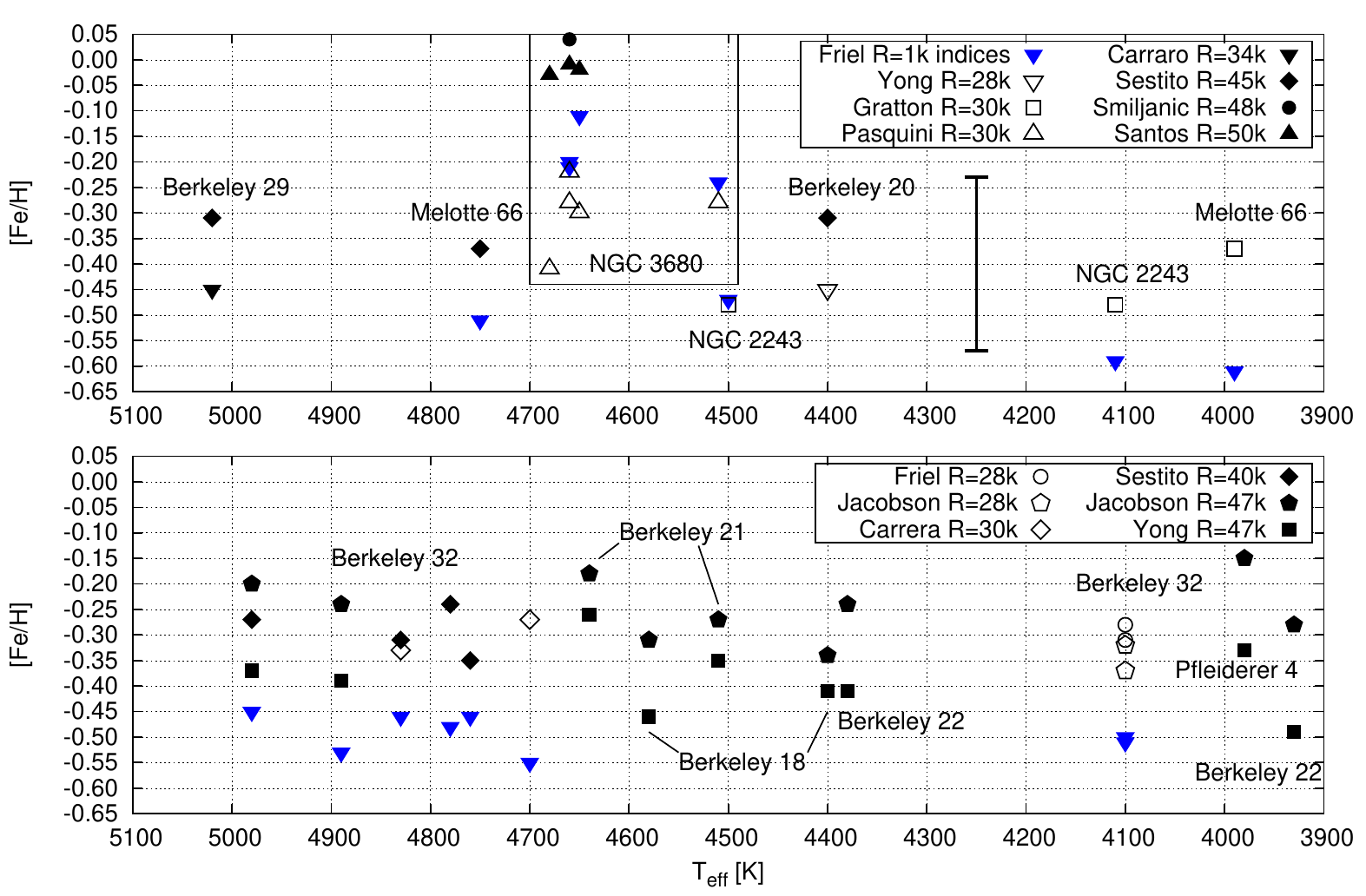}
   \caption{
   Same as Fig.~\ref{fig:M67giants}, for giant stars in ten different metal-poor clusters (five each in the upper and lower panels).
The metallicities are taken from the following publications:
\citet{1993AaA...267...75F,1994AaA...283..911G,2001AaA...374.1017P,2002AJ....124.2693F,2004AJ....128.1676C,2005AJ....130..597Y,2006AaA...458..121S,2008AaA...488..943S}; \citet[][Table~3]{2009AaA...493..309S}; \citet{2010AJ....139.1942F}; \citet{2011AaA...535A..30C}; \citet{2012AJ....144...95Y}; and \citet{2013AJ....145..107J}.
For star identifications and corresponding reference \Teff\ see Table~\ref{tab:metalpoor}.
The bar below the legend in the upper panel indicates the mean of the uncertainties for the individual [Fe/H] values quoted in the high-resolution publications (ranging from 0.04 to 0.3~dex). }
   \label{fig:metalpoor}
   \end{figure*}
%

\begin{table}
   \caption{Reference \Teff\ values and identifications of giant stars in ten different metal-poor clusters (see Fig.~\ref{fig:metalpoor}).}
\label{tab:metalpoor}
\centering
\begin{tabular}{lllrlr}
\hline\hline
\noalign{\smallskip}
\Teff    &  Cluster & \multicolumn{2}{c}{ID 1} & \multicolumn{2}{c}{ID 2}\\
\noalign{\smallskip}
\hline
\noalign{\smallskip}
3930 & Berkeley 22 & K    &  643 &      &     \\
3980 & Pfleiderer 4 & RGB & 1 & \multicolumn{2}{l}{J23505744+6220031\tablefootmark{$\dagger$}} \\
3990 & Melotte 66  & KJF  & 2236 &      & 4151 \\
4100 & Berkeley 32 & KM   &    2 & DBT  & 2689 \\
4100 & Berkeley 32 & KM   &    4 & DBT  & 1556 \\
4110 & NGC 2243    &      & 4209 & MMU  & 3633 \\
4380 & Berkeley 22 & K    &  414 &      &     \\
4400 & Berkeley 18 & K    & 1383 &      &     \\
4400 & Berkeley 20 & MPJF &    8 & SBR  & 1240 \\ 
4500 & NGC 2243    &      & 4110 & MMU  & 1313 \\
4510 & Berkeley 21 & TPM  &   51 &      & 415a\tablefootmark{$\ddagger$} \\
4510 & NGC 3680    & EGG  &   44 & AHTC & 1031 \\
4580 & Berkeley 18 & K    & 1163 &      &     \\
4640 & Berkeley 21 & TPM  &   50 &      &  50\tablefootmark{$\ddagger$} \\    
4650 & NGC 3680    & EGG  &   26 & AHTC & 3017 \\
4660 & NGC 3680    & EGG  &   13 & AHTC & 3003 \\
4660 & NGC 3680    & EGG  &   53 & KGP  & 1873 \\
4680 & NGC 3680    & EGG  &   41 & AHTC & 1050 \\
4700 & Berkeley 32 & KM   &   12 & DBT  & 1393 \\
4750 & Melotte 66  & KJF  & 1953 & SBR  & 1346 \\
4760 & Berkeley 32 & KM   &   19 & DBT  & 787 \\
4780 & Berkeley 32 & KM   &   27 & DBT  & 605 \\ 
4830 & Berkeley 32 & KM   &   17 & DBT  & 533 \\ 
4890 & Berkeley 32 & KM   &   16 & DBT  & 737 \\
4980 & Berkeley 32 & KM   &   18 & DBT  & 997 \\
5020 & Berkeley 29 & BHT  &  398 & FMP  & 948 \\
\noalign{\smallskip}
\hline\hline
\noalign{\smallskip}
\end{tabular}
\tablefoot{
Columns \emph{ID~1} and \emph{ID~2} give two alternative identifications, with acronyms used by the Simbad database, with some exceptions. \emph{ID~1} corresponds to the numbering system adopted by WEBDA, except for Berkeley~32 and Pfleiderer~4.
\tablefoottext{$\dagger$}{2MASS}
\tablefoottext{$\ddagger$}{Numbering by \citet{1979AJ.....84..204C}}
}
\end{table}

To evaluate the influence of spectrum quality and analysis methods on the metallicities of metal-poor clusters, we compared results from 13 publications since 1990 for giant stars appearing in ten different clusters. For these clusters, at least one determination gives a value of \FeH=$-0.3$~dex or lower.
We plot the metallicities determined for the individual stars in Fig.~\ref{fig:metalpoor} as a function of \Teff\ (see figure caption for references).
The reference \Teff\ values are taken from the works with highest resolution (or are averages from works with the same resolution).
Each \Teff\ value corresponds to one star, except for \Teff=4100~K in the lower panel and 4660~K in the upper panel, which correspond to two stars each.
The reference \Teff\ values and identifications of the stars are listed in Table~\ref{tab:metalpoor}. 
The reference \Teff\ values agree to within 200~K with those given in the individual publications.
The quoted uncertainties for the individual \FeH\ values range from 0.08 \mbox{\citep{2009AaA...493..309S}} to 0.3~dex \mbox{\citep{2001AaA...374.1017P,2013AJ....145..107J}}, and the mean of the uncertainties for high-resolution studies is 0.17~dex.

For these clusters, we can mainly compare the metallicities by \citet[][Melotte~66, Berkely~21]{1993AaA...267...75F} and \mbox{\citet{2002AJ....124.2693F}} based on spectroscopic indices to the high-resolution metallicities. The mean star-by-star difference for 14 stars (excluding two stars in \mbox{\citealt{2001AaA...374.1017P}}) is $-0.16\pm0.07$~dex, intermediate between the differences for M67 and NGC~6791 (both at higher metallicity).

Five of the stars appear in two or more high-resolution studies that are based on different observations: Berkeley~29 BHT~398, Berkeley~20 MPJF~8, and Berkeley~32 KM~17, as well as two stars in NGC~3680 (discussed below).
For the two stars in Berkeley~29 and Berkeley~20, the study by \mbox{\citet[][S/N=25--50 and 40--80, respectively]{2008AaA...488..943S}} resulted in higher metallicities (by 0.14~dex) than the studies by \mbox{\citet{2004AJ....128.1676C}} and \mbox{\citet[][S/N=70 and 56, respectively]{2005AJ....130..597Y}}, while the determinations for the star in Berkeley~32 by \mbox{\citet{2006AaA...458..121S}} and \mbox{\citet{2011AaA...535A..30C}} are in excellent agreement.
There are two other stars in Berkeley~32 (at \Teff=4100~K), for which \mbox{\citet{2010AJ....139.1942F}} and \mbox{\citet{2013AJ....145..107J}} analysed the same observed spectra at $R=30000$ and obtained consistent metallicities.
On the other hand, the lower panel of Fig.~\ref{fig:metalpoor} shows nine stars in five clusters, for which \mbox{\citet{2012AJ....144...95Y}} and \mbox{\citet{2013AJ....145..107J}} analysed the same spectra at $R=47000$. The derived metallicities are systematically different by 0.14~dex. Note that the largest differences of about 0.2~dex are seen for the two stars with \Teff$<$4000~K.

At this point, we refer to the extensive comparison work presented by \mbox{\citet[][Appendix]{2010AJ....139.1942F}} for Berkeley~32. They used the line lists and equivalent widths measured for nine stars by \mbox{\citet{2006AaA...458..121S}} and determined abundances and atmospheric parameters with their own methods. A detailed discussion of the possible effects of variations in individual analysis ingredients is given. For one star (KM~18), they independently measured equivalent widths in the spectrum used by \mbox{\citet{2006AaA...458..121S}} and found excellent agreement. On the other hand, using different selections of lines for the abundance determination (for a fixed set of atmospheric parameters) resulted in [Fe/H] differences of up to 0.15~dex.

For NGC~3680, the metallicities from high-resolution studies cluster around two significantly different values (no. 63 in Table~\ref{t:mean_paper}): \mbox{\citet{2009AaA...493..309S}} obtained a solar value (based on three giant and two dwarf stars), close to the results by \mbox{\citet{2009AaA...502..267S}} for one of the three giants and the results of \mbox{\citet{2008AaA...489..403P}} for the same two dwarfs at $R$=100000. On the other hand, \mbox{\citet{2001AaA...374.1017P}} determined low metallicities (\FeH$\approx -0.3$~dex) for six giant stars, which include the three stars studied by \mbox{\citet{2009AaA...493..309S}} and \mbox{\citet[][see Fig.~\ref{fig:metalpoor}]{2009AaA...502..267S}}.
First, we note that \mbox{\citet{2001AaA...374.1017P}} quoted metallicity errors of 0.25 to 0.35~dex, which are among the largest quoted in any publication. These errors were computed in an unconventional way -- standard deviations of line abundances (interpreted as random errors due to uncertainties in equivalent widths and atomic parameters) were added linearly and not in quadrature to metallicity errors due to uncertainties in atmospheric parameters (estimated to be 0.18~dex). The errors are thus probably overestimated compared with those of other authors. 
Second, for their final cluster metallicity, \mbox{\citet{2001AaA...374.1017P}} excluded the star EGG~41 with the lowest metallicity, because they did not trust the \Teff\ value, and the star EGG~34 (\FeH=$-$0.07~dex, not shown in Fig.~\ref{fig:metalpoor}), because they suspected it to be a binary star. This does not change the low cluster metallicity (\FeH=$-$0.27$\pm$0.03~dex). In addition, they added a +0.1~dex systematic error estimated from two Hyades stars, and quoted a final cluster metallicity of $-$0.17$\pm$0.12~dex -- closer to the other studies, but still significantly lower.
Because of these discrepancies, which indicate systematic errors in the determinations by \mbox{\citet{2001AaA...374.1017P}}, we decided to disregard these determinations for the computation of the average metallicity of NGC~3680 (Sect.~\ref{sect:final}).

\subsection{Peculiar case: Collinder 261}
\label{sect:peculiar}

\begin{table}
   \caption{Atmospheric parameters and identifications of giant stars in Collinder~261 (see Fig.~\ref{fig:Collinder261}).}
\label{tab:Collinder261}
\centering
\begin{tabular}{rrrllr}
\hline\hline
\noalign{\smallskip}
\Teff & \logg & \FeH & Reference & \multicolumn{2}{c}{ID} \\
\noalign{\smallskip}
\hline
\noalign{\smallskip}
 3950 & 0.5 & $-$0.01 & \citet{2007AJ....133.1161D} & PJM & 1871 \\
 3980 & 0.4 & $-$0.32 & \citet{2005AaA...441..131C} & PJM & 1871 \\
 4000 & 0.7 & $-$0.31 & \citet{2003AJ....126.2372F} & PJM & 1871 \\
 4180 & 1.6 & $-$0.08 & \citet{2005AaA...441..131C} & PJM & 2105 \\
 4300 & 1.8 & $-$0.02 & \citet{2007AJ....133.1161D} & PJM & 1485 \\
 4300 & 1.5 & $-$0.32 & \citet{2003AJ....126.2372F} & PJM & 2105 \\
 4340 & 1.8 & $-$0.06 & \citet{2005AaA...441..131C} & PJM & 1485 \\
 4350 & 1.7 &   +0.12 & \citet{2008AaA...488..943S} & SBR &    2 \\
 4400 & 2.1 & $-$0.03 & \citet{2007AJ....133.1161D} & PJM & 1481 \\
 4400 & 1.5 & $-$0.16 & \citet{2003AJ....126.2372F} & PJM & 1045 \\
 4450 & 1.8 & $-$0.01 & \citet{2007AJ....133.1161D} & PJM & 1045 \\
 4470 & 2.1 &   +0.01 & \citet{2005AaA...441..131C} & PJM & 1045 \\
 4490 & 2.2 & $-$0.11 & \citet{2003AJ....126.2372F} & PJM & 1080 \\
 4500 & 2.1 &   +0.00 & \citet{2005AaA...441..131C} & PJM & 1080 \\
 4500 & 2.1 &   +0.02 & \citet{2007AJ....133.1161D} & PJM & 1080 \\
 4500 & 2.0 &   +0.01 & \citet{2007AJ....133.1161D} & PJM &   27 \\
 4500 & 2.3 &   +0.16 & \citet{2008AaA...488..943S} & SBR &    6 \\
 4500 & 1.9 & $-$0.03 & \citet{2007AJ....133.1161D} & PJM &   29 \\
 4546 & 2.2 &   +0.18 & \citet{2008AaA...488..943S} & SBR &    7 \\
 4550 & 2.0 &   +0.00 & \citet{2007AJ....133.1161D} & PJM & 2001 \\
 4580 & 1.8 & $-$0.02 & \citet{2005AaA...441..131C} & PJM & 2001 \\
 4600 & 2.0 &   +0.00 & \citet{2007AJ....133.1161D} & PJM & 1526 \\
 4600 & 2.0 &   +0.14 & \citet{2008AaA...488..943S} & SBR &    5 \\
 4600 & 2.0 & $-$0.01 & \citet{2007AJ....133.1161D} & PJM & 1801 \\
 4650 & 2.3 & $-$0.01 & \citet{2007AJ....133.1161D} & PJM & 1472 \\
 4670 & 2.2 &   +0.09 & \citet{2008AaA...488..943S} & SBR &   11 \\
 4700 & 2.4 &   +0.20 & \citet{2008AaA...488..943S} & SBR &   10 \\
 4720 & 2.1 &   +0.04 & \citet{2008AaA...488..943S} & SBR &    9 \\
\noalign{\smallskip}
\hline\hline
\noalign{\smallskip}
\end{tabular}
\tablefoot{Column \emph{ID} gives the star identification, with acronyms used by the Simbad database.}
\end{table}

\begin{figure}
   \centering
   \includegraphics[width=\columnwidth,trim=10 50 10 50,clip]{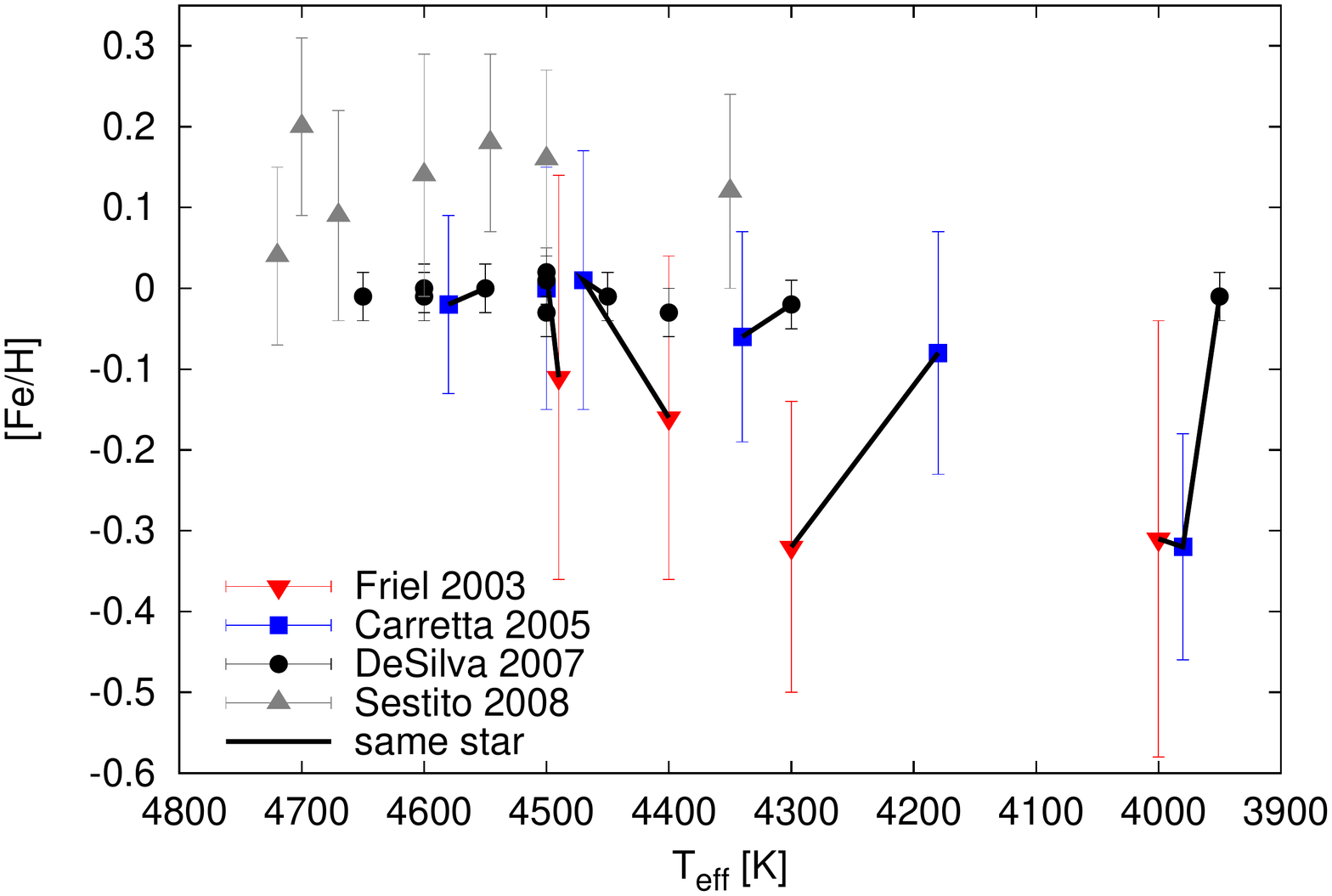}
   \caption{
   Metallicity versus \Teff\ for individual giant stars in Collinder~261.
   For star identifications, data, and references see Table~\ref{tab:Collinder261}. }
   \label{fig:Collinder261}
   \end{figure}

The cluster Collinder~261 has been studied in four different publications, all at high resolution and high S/N. The analyses resulted in four different cluster metallicities, from $-0.2$ to +0.1~dex in steps of 0.1~dex (no. 12 in Table~\ref{t:mean_paper}).
Figure~\ref{fig:Collinder261} shows the metallicity determinations for individual giant stars from all four works and their \FeH\ uncertainties. Six stars were studied by more than one author, and the data for these are connected with lines in the figure.
Star identifications, data, and references are given in Table~\ref{tab:Collinder261}.
All four works are based on similar-quality data from three different instruments (S/N$\approx$100, $R\approx 45000$, except for \mbox{\citealt{2003AJ....126.2372F}} with $R\approx 25000$).
All four works are based on an equivalent-width analysis, and three of them used the same code for computing the model equivalent widths \mbox{\citep[MOOG,][]{1973PhDT.......180S}}. They derived the stellar atmospheric parameters in the same way, forcing excitation and ionization equilibrium on the line abundances.
Thus, the diverging results obtained by these authors could be due to differences in the model atmospheres, spectral-line selection, and atomic data. Additional probable sources for discrepancies are continuum tracing and equivalent-width measurement \mbox{\citep{2008AaA...488..943S}}.

In three of the publications, the model atmospheres were those of R.~Kurucz, although different versions were used: \mbox{\citet{2005AaA...441..131C}} and \mbox{\citet{2008AaA...488..943S}} used models from the grid on CDROM \mbox{\citep{1993KurCD..13.....K}}, while \mbox{\citet{2007AJ....133.1161D}} interpolated in the grid published by \mbox{\citep{1997AaA...318..841C}}. \mbox{\citet{2003AJ....126.2372F}} interpolated in the grid of \mbox{\citet{1976AaAS...23...37B}}.
In Table~\ref{tab:Col261lines}, we list the number of iron lines used per star and the corresponding wavelength ranges. These properties of the line lists are very similar, except for that of \mbox{\citet{2007AJ....133.1161D}}, which includes bluer wavelengths.
All four works derived metallicities with respect to a reference object.
\mbox{\citet{2003AJ....126.2372F}} used oscillator-strength values derived from an Arcturus spectrum.
\mbox{\citet{2005AaA...441..131C}} and \mbox{\citet{2008AaA...488..943S}} quoted their Fe abundances relative to abundances derived from a solar analysis using equivalent widths measured from high-resolution solar atlases.
The homogeneity of the abundances by \mbox{\citet{2007AJ....133.1161D}} within the cluster was achieved by a differential line-by-line abundance analysis relative to one of the cluster stars, whereas the abundance zero-point was given by the adopted solar abundance from \mbox{\citet{1992AJ....104.2121S}}. 
These analysis approaches should minimize the uncertainties caused by atomic line data, although to different degrees.
The determination of gravity could be a major source for the differences, because it relies on ionization balance between abundances from Fe~I and Fe~II lines.
Since Fe~II lines are scarce in spectra of cool stars, the derived gravity critically depends on the set of Fe~II lines used and the accuracy of their equivalent widths.
However, for only two of the six stars in common, there are significant differences in gravity (PJM~1871 -- the coolest star, and PJM~1045 at 4400--4500~K, see Table~\ref{tab:Collinder261}), and the differences in gravity do not seem to be correlated with differences in abundances.

\begin{table}
   \caption{Abundance analyses of Collinder~261 stars -- approximate number of Fe~I and Fe~II lines used per star, and corresponding wavelength ranges.}
\label{tab:Col261lines}
\centering
\begin{tabular}{lrrrr}
\hline\hline
\noalign{\smallskip}
          & \multicolumn{2}{c}{Fe~I} & \multicolumn{2}{c}{Fe~II} \\
Reference & n & $\lambda$ [nm] & n & $\lambda$ [nm] \\
\noalign{\smallskip}
\hline
\noalign{\smallskip}
\citet{2003AJ....126.2372F} &  40 & 538 -- 785 &  8 & 541 -- 652 \\
\citet{2005AaA...441..131C} & 100 & 550 -- 700 & 10 & 550 -- 700 \\
\citet{2007AJ....133.1161D} &  60 & 422 -- 620 & 12 & 449 -- 615 \\
\citet{2008AaA...488..943S} & 100 & 550 -- 681 & 13 & 553 -- 652 \\
\noalign{\smallskip}
\hline\hline
\end{tabular}
\end{table}
%

\mbox{\citet{2005AaA...441..131C}} provided a detailed comparison of their work with \mbox{\citet{2003AJ....126.2372F}}. They showed that the equivalent widths of \mbox{\citet{2003AJ....126.2372F}} are systematically smaller for stronger lines ($\gtrsim$80~m\AA) for the lines in common (among which there are 24 Fe~I lines). 
This could in part explain the lower metallicities. On the other hand, \mbox{\citet{2005AaA...441..131C}} used the \mbox{\citet{2003AJ....126.2372F}} equivalent widths and their own methods for the star with the largest abundance difference (PJM~2105), and derived an \FeH\ value close to that obtained from their own data. They ascribed this to their different approach in estimating the microturbulent velocity (based on computed instead of observed equivalent widths).
For the test star, the microturbulence derived by \mbox{\citet{2005AaA...441..131C}} is indeed 0.25~kms$^{-1}$ lower. However, for two other stars with (smaller) abundance differences, the microturbulence is about 0.1~kms$^{-1}$ higher.

\mbox{\citet{2007AJ....133.1161D}} compared their results with those of \mbox{\citet{2005AaA...441..131C}}, and noted that they are similar except for the coolest star with lowest gravity (PJM~1871).
They also performed a test on three stars, using two different types of model atmospheres (Kurucz and MARCS, \mbox{\citealt{1997AaA...318..521A}}), and found that the differences in Fe abundance were lower than 0.03~dex.

Collinder~261 is a rather old OC, located in the inner disk, which makes it an interesting object for studying the evolution of metallicity distributions in the Galactic disk. It is included in the list of standard clusters proposed by \mbox{\citet{2006MNRAS.371.1641P}}, with an age of 8.8~Gyr and a distance of 2.2~kpc.
There is no photometric metallicity determination available, that is, it is not included in Paper~I.
Therefore, we used the method of \mbox{\citet{2010AaA...514A..81P}} (see also Sect.~\ref{sect:final}) to estimate its metallicity using $B, V, I$ photometry, published by \mbox{\citet{1996MNRAS.283...66G}} and \mbox{\citet[][$V$ and $I$ only]{1994AJ....107.1079P}}. The \mbox{\citet{1994AJ....107.1079P}} photometry was brought onto the scale by \mbox{\citet{1996MNRAS.283...66G}} by applying corrections of +0.031 and +0.033~mag for $V$ and $V-I$, respectively. An average and standard deviation of the photometry was calculated. We selected only stars available in both studies with errors $<$~0.05~mag, covering the whole cluster. The colour-colour and colour-magnitude diagrams were examined to select most probable main-sequence cluster members.

The method is based on grids of evolutionary models, originally for [Fe/H] from $-0.7$ to $+0.4$ and ages up to 4~Gyr, which were extended to older ages using Geneva isochrones \mbox{\citep{2001A&A...366..538L}}. To transform $V-I$ into effective temperatures (in addition to $B-V$), the empirical relation for A to K dwarfs by \mbox{\citet{1998A&A...333..231B}} as well as their colour-excess ratios (including colour terms) were used.To transform the photometry into the $T_\mathrm{eff}-\mathrm{log}(L/L_{\odot})$ plane, the cluster parameters by \mbox{\citet{2006MNRAS.371.1641P}} were adopted. Since only $E(B-V)$ is tabulated, the temperatures deduced from $V-I$ were scaled to the $B-V$ results by applying an offset of $-$0.04~mag to the transformed reddening value.
This offset can either be due to a small error in the calibration or to an abnormal reddening law in this direction. The final temperatures from both indices agree within $<$\,2\%, and were averaged. All other steps of the method were applied as given in \mbox{\citet{2010AaA...514A..81P}}.

\begin{figure}
\includegraphics[width=\columnwidth,trim=10 10 5 5,clip]{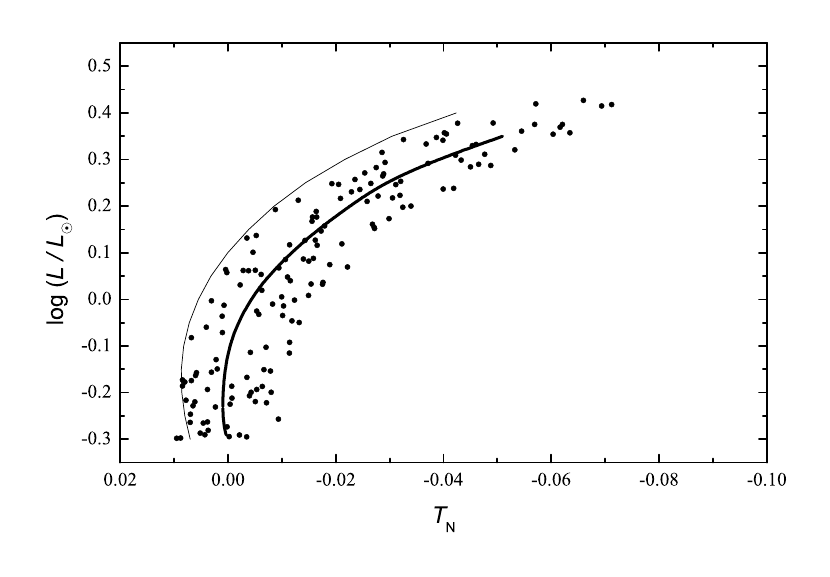}
\caption{Theoretical HR-diagram illustrating the photometric metallicity determination for Collinder~261 using the method by \citet{2010AaA...514A..81P}. 
Logarithmic luminosity is plotted versus normalized logarithmic effective temperature, $T_{\rm N}=\log$\Teff$-\log$\Teff(ZAMS), where \Teff(ZAMS) is the \Teff\ of a zero-age main-sequence with solar composition for a given luminosity (see Table~4 of \citealt{2010AaA...514A..81P}).}
Using the parameters by \citet{2006MNRAS.371.1641P}, the metallicity was determined as $Z$=0.021 ([Fe/H]$\approx$0.05~dex, isochrone shown as thick line). In addition, the isochrone for $Z$=0.012 ([Fe/H]$\approx-$0.2, thin line) is shown, representing the high-resolution metallicity value of \citet{2003AJ....126.2372F}.
\label{f:cr261hrd}
\end{figure}

In Fig.~\ref{f:cr261hrd}, the best-fit isochrone for the given parameters can be found, which results in a metallicity estimate of $Z$=0.021 (heavy-element-mass fraction, corresponding to [Fe/H]$\approx$0.05~dex for $Z_\odot$=0.019 and a helium-mass fraction $Y=0.23+2.25Z$). This supports the spectroscopic results around solar metallicity or higher.
In addition, the low-resolution result by \mbox{\citet{2002AJ....124.2693F}} with [Fe/H]=$-$0.16~dex would support solar metallicity, when applying an ad-hoc correction of +0.2~dex, as suggested by comparison with high-resolution results (see e.g. Sect.~\ref{sect:mean}).

In conclusion, we decided to disregard the determinations by \mbox{\citet{2003AJ....126.2372F}} for the computation of the average metallicity of Collinder~261 (Sect.~\ref{sect:final}) because of the lower resolution and the inconsistencies with the other three studies.

\subsection{Other cases with high dispersion or large discrepancies}
\label{sect:other}

\paragraph{Melotte 20:}
The highest dispersion in the high-resolution sample occurs for Melotte~20 ($\alpha$~Per cluster), studied by \mbox{\citet[][see Table~\ref{t:mean_paper}]{1996AJ....111..424G}} -- 0.26~dex for three stars, one supergiant and two dwarfs.
For one of the dwarf stars (HE~490), their result is close to the only other high-resolution study for four dwarf stars in this cluster \mbox{\citep{1990ApJ...351..467B}}, and to metallicity estimates from lower-resolution spectroscopy and photometry.
The metallicity of the second dwarf star (HE~767) is about 0.2~dex higher. \mbox{\citet{1996AJ....111..424G}} discussed several possible explanations for this discrepancy, but did not find supporting evidence for any of them and concluded that the problem remains unresolved.
The metallicity of the supergiant ($\alpha$~Per, \logg=1.2) is about 0.3~dex lower than that of HE~490. The authors ascribed this discrepancy to non-LTE effects, based on a contemporary estimation of these effects for metal-poor stars with similar \Teff, but slightly larger \logg\ \mbox{\citep{1996ApJS..103..183L}}.
We note that the \mbox{\citet{1996AJ....111..424G}} analysis used stellar-atmosphere models obtained from R.~Kurucz in 1992, which assume plane-parallel geometry. However, the extended atmospheres of supergiants are more accurately described by spherical geometry. According to \mbox{\citet{Heit:06}}, the inappropriate geometry leads to an overestimation of Fe abundances by $\approx$0.05~dex for the stellar parameters and Fe lines used by \mbox{\citet{1996AJ....111..424G}}, that is, in the opposite direction of the non-LTE effect.

\paragraph{IC~4651:}
This cluster appears in four high-resolution studies (no. 16 in Table~\ref{t:mean_paper}). In the study by \mbox{\citet{2004AaA...422..951C}}, the large dispersion of 0.19~dex obtained for the average metallicity of five stars agrees with the individual errors quoted by the authors. However, the large dispersion is due to one star, a cool giant near the RGB tip (IC~4651 56, \logg=0.3). The metallicity of this star is about 0.4~dex lower than that of the other four stars, which are red clump stars. Again, the authors ascribed the discrepancy to non-LTE effects, supported by the large difference between spectroscopic and photometric (evolutionary) gravities determined for the RGB tip star. For this type of star, geometry only affects abundances derived from Fe~II lines, which may be underestimated by $\approx$0.05~dex \mbox{\citep{Heit:06}}. The mean cluster metallicity quoted by the authors (0.11$\pm$0.01~dex) is based on the four red clump stars. It has the smallest dispersion of the four studies and agrees perfectly with the others.


\paragraph{Collinder~121:}
Only one star in this cluster, the supergiant HD~50877, has been studied by two authors.
\mbox{\citet{1998AaA...338..623M}} quoted \FeH=+0.25, and \mbox{\citet{2007AaA...475.1003H}} obtained \FeH=$-$0.32. They also quoted substantially different \Teff\ and \logg\ values (3200~K/0.0 and 3900~K/0.65, respectively). As mentioned above, non-LTE effects are suspected to occur in the atmospheres of supergiants, and therefore these stars are poorly suited to estimate the metallicity of an OC.


\paragraph{NGC~2141:}
Two different stars were studied in three publications (one each by \mbox{\citealt{2005AJ....130..597Y}} and \mbox{\citealt{2009AJ....137.4753J}}, and both by \mbox{\citealt{2013AJ....145..107J}}). Both stars are bright giants (\logg=1.2). The first two works resulted in rather different metallicities ($-0.18\pm0.15$~dex and +0.00$\pm$0.16~dex, respectively). Although these values agree within the quoted errors, the large errors and the large discrepancy point to considerable uncertainties inherent in the atmospheric modelling of these stars.
\mbox{\citet[][Sect. 5.1.3]{2009AJ....137.4753J}} investigated this problem in more detail. They obtained the spectrum used by \mbox{\citet{2005AJ....130..597Y}} taken with the same instrument and setup, and two more spectra of the same stars with higher resolution, but lower S/N used by \mbox{\citet{2004PhDT........36B}}, who had derived an even lower metallicity for the two stars. Analysing these spectra from scratch in the same way as their own observations, they arrived at a consistent metallicity close to solar. They demonstrated that different values for the microturbulence parameter are a major source for the abundance differences. Systematic differences in measured equivalent widths and different sets of spectral lines may contribute as well.
\mbox{\citet{2013AJ....145..107J}} reanalysed the same spectra as were used in \mbox{\citet{2005AJ....130..597Y}} and \mbox{\citet{2009AJ....137.4753J}} with a new version of the radiative transfer code and more recent, spherical stellar atmosphere models than \mbox{\citet{2009AJ....137.4753J}}. They obtained consistent metallicities of $-0.09\pm0.18$~dex for the two stars, and for the Fe~I and the Fe~II line lists, that is, in between the previous discrepant determinations.


\begin{table*}
   \caption{Abundance analyses of NGC~2632 stars (Praesepe) -- mean [Fe/H] values and standard deviation, spectrum quality, approximate number of Fe~I and Fe~II lines used, and corresponding wavelength ranges.}
\label{t:praesepe}
\centering
\begin{tabular}{llrrrrrr}
\hline\hline
\noalign{\smallskip}
       &           &   &     & \multicolumn{2}{c}{Fe~I} & \multicolumn{2}{c}{Fe~II} \\

 [Fe/H] & Reference & R & S/N & n & $\lambda$ [\AA] & n & $\lambda$ [\AA] \\
\noalign{\smallskip}
\hline
\noalign{\smallskip}
\multicolumn{7}{l}{dwarfs} \\
$+$0.04 $\pm$ 0.03 & \citet{1992ApJ...387..170F}    &  28k & 100--200    & 7/14 & 7015--7205 &    &            \\
$+$0.11 $\pm$ 0.00 & \citet{2007ApJ...655..233A}    &  55k & $>$100      &   15 & 5300--6200 &  9 & 4490--6150 \\ 
$+$0.27 $\pm$ 0.04 & \citet{2008AaA...489..403P}    & 100k & $\approx$80 &   60 & 4800--6800 & 10 & 4800--6800 \\
\multicolumn{7}{l}{giants} \\
$+$0.15 $\pm$ 0.05 & \citet{2011AaA...535A..30C}    &  30k & 150--215    &  177 & 5055--8945 &  9 & 5991--7711 \\ 
\noalign{\smallskip}
\hline\hline
\end{tabular}
\end{table*}

\paragraph{NGC~2632:}
For Praesepe, metallicity determinations are available for 15 different stars (three giants and twelve dwarfs) in four publications (no. 58 in Table~\ref{t:mean_paper}). In each paper, a different set of stars was analysed, which makes a direct comparison impossible. The results in each of the three publications using dwarf stars show small dispersions for their respective stellar samples (at most 0.04~dex for two to six stars), but the mean metallicities are significantly different from each other (by 0.07 to 0.23~dex). The discrepancies might be due, apart from the different objects studied, to the different spectroscopic material, and the different selection of Fe lines used.  In particular, the wavelength regions and spectral resolutions vary between the three works, while the S/Ns are similar (see Table~\ref{t:praesepe}).
It is curious that the study based on the spectra with the highest resolution quotes the most discrepant value.
The metallicity derived by \mbox{\citet{2007ApJ...655..233A}} for dwarf stars agrees best with the value obtained for giants by \mbox{\citet{2011AaA...535A..30C}}.


\subsection{Importance of spectrum quality for mean cluster metallicity}
\label{sect:mean}

   \begin{figure*}
   \centering
   \includegraphics[width=\textwidth,trim=10 40 10 50,clip]{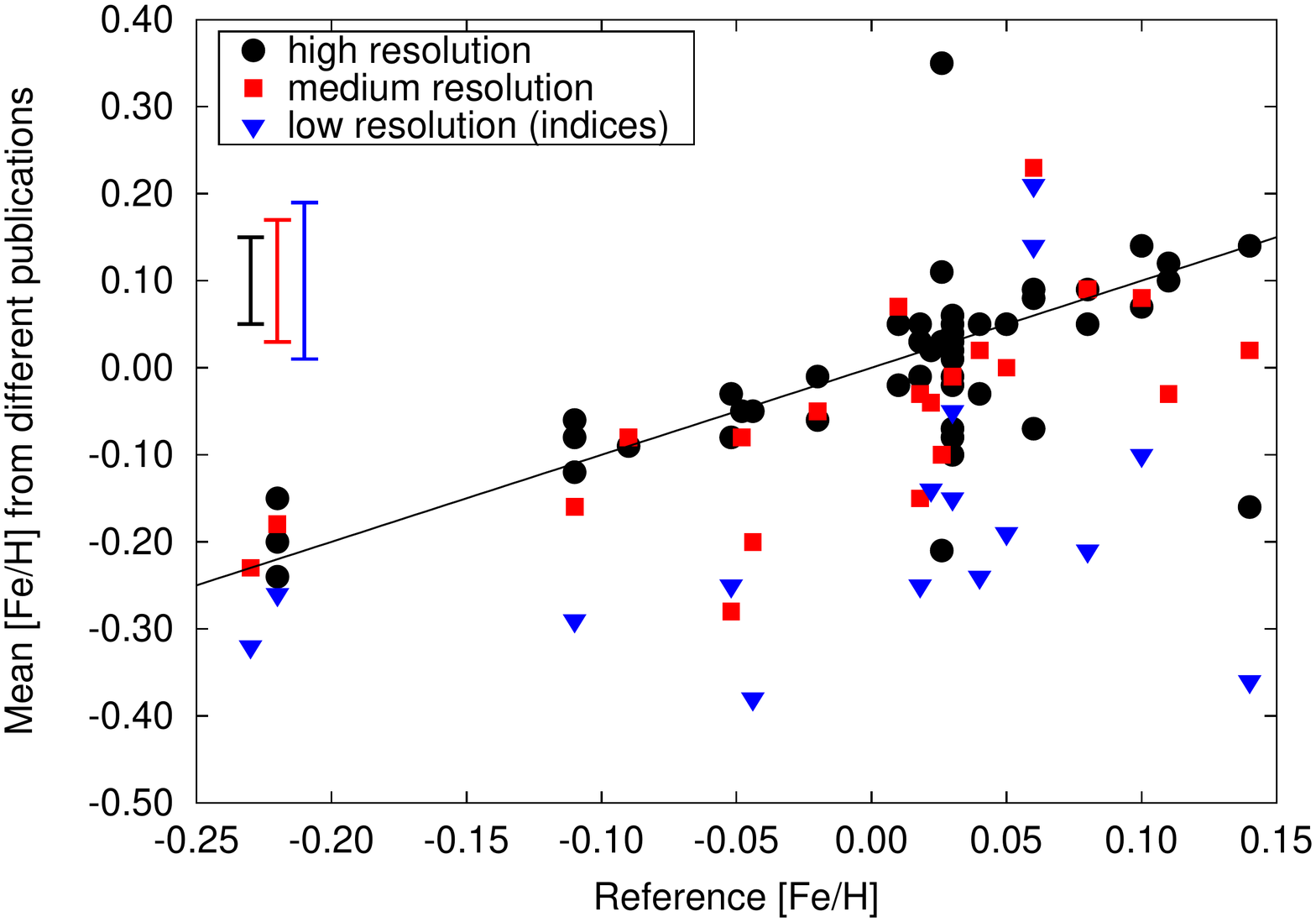}
   \caption{
   Mean cluster metallicity determined by different authors from spectra with different resolution versus reference \FeH. High resolution: $R\ge25000$, medium resolution: $R\ge13000$, low resolution: $R\approx 1000$, spectroscopic indices. All spectra have high S/N. The reference [Fe/H] values are given in Table~\ref{tab:mean1}. The solid line is the one-to-one relation. The bars below the legend represent the average standard deviations of the cluster means for high, medium, and low resolution from left to right, respectively. Metallicities and references for medium- and low-resolution studies are given in Table~\ref{tab:mean1}. Metallicities and references for high-resolution studies are given in Table~\ref{t:mean_paper}.}
   \label{fig:mean1}
   \end{figure*}
%

\begin{table*}
\caption{Metallicities and references for medium- and low-resolution studies for clusters shown in Fig.~\ref{fig:mean1}.}
\label{tab:mean1}
\centering
\begin{tabular}{lrrrrrll}
\hline\hline
\noalign{\smallskip}
Cluster ID & ref. \FeH & mean \FeH & std. dev. & resolution & \# & star type & reference \\
\noalign{\smallskip}
\hline
\noalign{\smallskip}
NGC2243            & $-$0.480 & $-$0.42 & 0.05 & 18000$-$20000 & 10 & giant & \citet{2011AJ....141...58J} \cr
NGC2243            & $-$0.480 & $-$0.49 & 0.05 & 1250 &  9 & giant & \citet{2002AJ....124.2693F} \cr
NGC2204            & $-$0.230 & $-$0.23 & 0.04 & 20000 & 13 & giant & \citet{2011AJ....141...58J} \cr
NGC2204            & $-$0.230 & $-$0.32 & 0.10 & 1250 & 12 & giant & \citet{2002AJ....124.2693F} \cr
Berkeley39         & $-$0.220 & $-$0.18 & 0.06 & 22500 & 21 & giant & \citet{2012AaA...548A.122B} \cr
Berkeley39         & $-$0.220 & $-$0.26 & 0.09 & 1250 & 14 & giant & \citet{2002AJ....124.2693F} \cr
NGC1817            & $-$0.110 & $-$0.16 & 0.03 & 18000/21000 & 28 & giant & \citet{2011AJ....142...59J} \cr
NGC1817            & $-$0.110 & $-$0.29 & 0.05 & 1250 &  3 & giant & \citet{2002AJ....124.2693F} \cr
NGC2194            & $-$0.090 & $-$0.08 & 0.08 & 18000 &  6 & giant & \citet{2011AJ....142...59J} \cr
NGC2158            & $-$0.052 & $-$0.25 & 0.09 & 1250 &  7 & giant & \citet{2002AJ....124.2693F} \cr
NGC2158            & $-$0.052 & $-$0.28 & 0.05 & 14500/21000 & 15 & giant & \citet{2011AJ....142...59J} \cr
NGC2355            & $-$0.048 & $-$0.08 & 0.08 & 18000 &  5 & giant & \citet{2011AJ....142...59J} \cr
NGC2420            & $-$0.044 & $-$0.20 & 0.06 & 18000 &  9 & giant & \citet{2011AJ....142...59J} \cr
NGC2420            & $-$0.044 & $-$0.38 & 0.07 & 1250 & 20 & giant & \citet{2002AJ....124.2693F} \cr
IC2602             & $-$0.020 & $-$0.05 & 0.04 & 18000 &  6 & dwarf & \citet{2001AaA...372..862R} \cr
NGC3532            & $+$0.010 & $+$0.07 & 0.06 & 18000 &  5 & giant & \citet{1994ApJS...91..309L} \cr
IC4756             & $+$0.018 & $-$0.03 & 0.05 & 18000 &  4 & giant & \citet{1994ApJS...91..309L} \cr
IC4756             & $+$0.018 & $-$0.15 & 0.04 & 15000 &  6 & giant & \citet{2007AJ....134.1216J} \cr
IC4756             & $+$0.018 & $-$0.25 & 0.08 & 1250 &  8 & giant & \citet{2002AJ....124.2693F} \cr
NGC1245            & $+$0.022 & $-$0.04 & 0.05 & 18000 & 13 & giant & \citet{2011AJ....142...59J} \cr
NGC1245            & $+$0.022 & $-$0.14 & 0.09 & 1150 &  7 & clump & \citet{2005AJ....130.1916M} \cr
NGC6633            & $+$0.026 & $-$0.10 & 0.02 & 13000 & 10 & dwarf & \citet{2002MNRAS.336.1109J} \cr
NGC2682            & $+$0.030 & $-$0.01 & 0.05 & 14500/18000 & 19 & giant & \citet{2011AJ....142...59J} \cr
NGC2682            & $+$0.030 & $-$0.05 & 0.04 & 1150 &  5 & clump & \citet{2005AJ....130.1916M} \cr
NGC2682            & $+$0.030 & $-$0.15 & 0.05 & 1250 & 25 & giant & \citet{2002AJ....124.2693F} \cr
NGC7789            & $+$0.040 & $+$0.02 & 0.04 & 18000 & 28 & giant & \citet{2011AJ....142...59J} \cr
NGC7789            & $+$0.040 & $-$0.24 & 0.09 & 1250 & 57 & giant & \citet{2002AJ....124.2693F} \cr
NGC6939            & $+$0.050 & $-$0.19 & 0.09 & 1250 &  4 & giant & \citet{2002AJ....124.2693F} \cr
NGC6939            & $+$0.050 & $+$0.00 & 0.10 & 15000 &  8 & giant & \citet{2007AJ....134.1216J} \cr
NGC6705            & $+$0.060 & $+$0.14 & 0.16 & 1150 &  4 & clump & \citet{2005AJ....130.1916M} \cr
NGC6705            & $+$0.060 & $+$0.21 & 0.09 & 1200/2000 &  6 & giant & \citet{1993PASP..105.1253T} \cr
NGC6705            & $+$0.060 & $+$0.23 & 0.13 & 24000 &  6 & giant & \citet{2000PASP..112.1081G} \cr
NGC5822            & $+$0.080 & $-$0.21 & 0.10 & 1250 &  3 & giant & \citet{2002AJ....124.2693F} \cr
NGC5822            & $+$0.080 & $+$0.09 &      & 18000 &  1 & giant & \citet{1994ApJS...91..309L} \cr
NGC7142            & $+$0.100 & $-$0.10 & 0.10 & 1250 & 12 & giant & \citet{2002AJ....124.2693F} \cr
NGC7142            & $+$0.100 & $+$0.08 & 0.06 & 15000 &  6 & giant & \citet{2007AJ....134.1216J} \cr
NGC188             & $+$0.110 & $-$0.03 & 0.04 & 18000 & 27 & giant & \citet{2011AJ....142...59J} \cr
NGC2112            & $+$0.140 & $-$0.36 &      & 1250 &  1 & giant & \citet{2002AJ....124.2693F} \cr
NGC2112            & $+$0.140 & $+$0.02 &      & 16000 &  1 & giant & \citet{1996AJ....112.1551B} \cr
NGC6791            & $+$0.420 & $+$0.11 & 0.10 & 1250 & 39 & giant & \citet{2002AJ....124.2693F} \cr
NGC6791            & $+$0.420 & $+$0.32 & 0.02 & $\approx$1000 & 14 & giant & \citet{2003PASP..115...96W} \cr
NGC6791            & $+$0.420 & $+$0.42 & 0.05 & 15000 & 16 & giant & \citet{2012ApJ...756L..40G} \cr

\noalign{\smallskip}
\hline\hline
\noalign{\smallskip}
\end{tabular}
\tablefoot{
The second column lists the reference metallicity. The reference [Fe/H] values are the weighted means of high-resolution determinations for all stars in each cluster (cf. Table~\ref{tab:appendix_members}; IC~4756, NGC~1245, and NGC~6633, as well as NGC~2158, NGC~2355, and NGC~2420 were shifted for better visibility) or the medium-resolution determination for NGC~2204.
}
\end{table*}
%

We now turn from comparing determinations for individual stars to a comparison of mean cluster metallicities based on low-, medium- and high-resolution spectra, \emph{all with high S/N}.
We started from a list of 33 clusters studied at medium resolution ($13000\le R<25000$) and searched for low- or high-resolution determinations for the same clusters.
In total, we found multiple-resolution determinations for 23 clusters.
Nineteen of these have determinations from low-resolution spectra (spectroscopic indices) in four publications. The medium-resolution determinations are taken from ten publications.
The low- and medium-resolution references are listed in Table~\ref{tab:mean1}.

The cluster Tombaugh~2 is not included in Table~\ref{tab:mean1}. It is part of the \mbox{\citet{2002AJ....124.2693F}} sample, and has been studied at medium resolution by \mbox{\citet{2008MNRAS.391...39F}} and \mbox{\citet{2010AaA...509A.102V}}. These two authors analysed 14 and 13 radial velocity members, respectively, using the same instrument, but different wavelength regions. \mbox{\citet{2008MNRAS.391...39F}} obtained two different mean metallicities (about solar and $-0.3$~dex) for the two halves of their sample, while \mbox{\citet{2010AaA...509A.102V}} arrived at a consistent metallicity of about $-0.3$~dex for all of their stars. The possible reasons for the different results are discussed at length in \mbox{\citet{2010AaA...509A.102V}}. We add that of the five stars in common between the two studies, one has the same atmospheric parameters and metallicity, while three have different parameters and metallicities, and one has different parameters but the same metallicity. We note in particular that the metallicity difference increases with the difference in microturbulence parameter between the two works.
Two additional radial-velocity members of Tombaugh~2 have been analysed by \mbox{\citet{1996AJ....112.1551B}} at high resolution. However, no atmospheric parameters are given, and for the metallicity only the absolute value is quoted. Adopting the reference solar abundance for Fe~I given in \mbox{\citet{1992AJ....104.1818B}}, who used the same analysis methods, the metallicities of the two stars would be $-0.6$ and $-0.8$~dex.
We regard the status of this cluster as inconclusive. 

All but one cluster are included in the high-resolution sample (31 publications, see Table~\ref{t:mean_paper}).
Most of them have several different determinations that agree very well with each other.
In Fig.~\ref{fig:mean1}, we plot the mean cluster metallicities from each publication as a function of reference \FeH\ (between $-0.25$ and +0.15~dex), as given in Table~\ref{tab:mean1}.
Two clusters at the two metallicity extremes lie outside of this range (NGC~2243 and NGC~6791).

The metallicities based on indices by \mbox{\citet{2002AJ....124.2693F}} are lower by up to 0.5~dex than the medium- and high-resolution metallicities (the differences increase with metallicity).
This discrepancy has previously been noted for individual stars in representative clusters in Sects.~\ref{sect:m67} to \ref{sect:metalpoor},
and by \mbox{\citet{2010AaA...511A..56P}} for 28 OCs in the \mbox{\citet{2002AJ....124.2693F}} sample, which also have $R>$15000 determinations \mbox{\citep[see also][Sect.~6.1.3]{2010AJ....139.1942F}}.
A systematic difference of about 0.2~dex is also seen between the \mbox{\citet{2002AJ....124.2693F}} OC metallicities and metallicities compiled for Cepheids at similar Galactocentric distances \mbox{\citep[see Fig.~4 in][]{2009AaA...504...81P}}.
On the other hand, for NGC~6705 at \FeH=0.06~dex, both the \mbox{\citet{1993PASP..105.1253T}} and \mbox{\citet{2005AJ....130.1916M}} indices result in higher metallicities than at high resolution, while for NGC~1245 (0.02~dex), NGC~2682 (0.03~dex), and NGC~6791 (0.42~dex), the determinations by \mbox{\citet{2005AJ....130.1916M}} or \mbox{\citet{2003PASP..115...96W}} are about 0.1~dex lower than at high resolution.

The medium-resolution results agree in general very well with the high-resolution ones, except for IC~4756, NGC~188, NGC~2158, NGC~2420, and NGC~6705.
For IC~4756 at 0.018~dex the work of \mbox{\citet{2007AJ....134.1216J}} gives a 0.15~dex lower metallicity value than the other medium-resolution publication and the three high-resolution publications.
For NGC~188, NGC~2158, and NGC~2420 the medium-resolution values by \mbox{\citet{2011AJ....142...59J}} are 0.1 to 0.2~dex lower than the high-resolution determinations.
For NGC~6705 at 0.06~dex, the medium-resolution value by \mbox{\citet{2000PASP..112.1081G}} is 0.15~dex higher than two of the high-resolution values, while the high-resolution value of \mbox{\citet{2012AaA...538A.151S}} is lower than the other two by a similar amount, and all mean \FeH\ values agree within the standard deviations.


NGC~2112 (reference \FeH=0.14~dex) has two discrepant high-resolution determinations. This old anticentre OC is useful for determining the radial metallicity gradient of the Galactic disk. \mbox{\citet{1996AJ....112.1551B}} found it to be mildly deficient, \FeH=$-0.16\pm0.25$~dex (only one star was observed at high resolution), while \mbox{\citet{2008MNRAS.386.1625C}} found it to be slightly supersolar, \FeH=+0.14~dex, close to the Hyades value. \mbox{\citet{1996AJ....112.1551B}} observed a cool red giant, while \mbox{\citet{2008MNRAS.386.1625C}} observed one dwarf, one clump giant, and one F~giant with a small dispersion of 0.03~dex.

NGC~6633 (reference \FeH=0.026~dex) has four discrepant high-resolution determinations.
The closest \FeH\ values, +0.03~dex for six measurements of three stars, and +0.11~dex for one star, are obtained by \mbox{\citet{2009AaA...493..309S}} and \mbox{\citet{2009AaA...502..267S}}, respectively.
\mbox{\citet{2005ApJS..159..141V}} observed the same star as \mbox{\citet{2009AaA...502..267S}}, NGC~6633 100 = HD~170174, which is also in common with \mbox{\citet{2009AaA...493..309S}}, but derived the significantly higher value of +0.35~dex.
The temperature adopted by \mbox{\citet{2005ApJS..159..141V}} is substantially hotter (\Teff=5245~K, while \mbox{\citealt{2009AaA...502..267S}} adopted \Teff=5015~K, and \mbox{\citealt{2009AaA...493..309S}} \Teff=4980~K). We conclude that the different temperature scale employed by \mbox{\mbox{\citet{2005ApJS..159..141V}}} is the cause for the discrepant metallicity determination. Therefore, we discarded this determination when we calculated the final mean cluster metallicity (Sect.~\ref{sect:final}). The fourth metallicity determination by \mbox{\citet{2005MNRAS.363L..81A}} is significantly lower than the others ($-0.21$~dex) and is based on two F~dwarfs. The most metal-poor one (JEF~1), with \FeH=$-$0.31~dex is a moderate rotator with $v$sin$i$=19~km~s$^{-1}$ and a high \Teff = 6870~K, while the cooler F~dwarf (HJT~1251 = JEF~16) has a value of \FeH=$-$0.15~dex, which agrees better with the other determinations.

In summary, at resolutions higher than $\sim$10000, cluster metallicities may have small or large dispersions regardless of the resolution value. Other factors such as analysis method, temperature scale, or properties of the sample stars seem to play a larger role for the reliability of the metallicity than spectral resolution.

%

Finally, we assessed the impact of S/N by comparing mean cluster metallicities determined by different authors from spectra with high ($>50$) or low S/N, \emph{all of them with high resolution}. For six clusters we found metallicity determinations from both high-S/N spectra (21 publications) and low-S/N spectra (8 publications). The clusters, metallicities, and references for low-S/N determinations are given in Table~\ref{tab:mean2}. Metallicities and references for high-S/N determinations can be found in Table~\ref{t:mean_paper}.
For NGC~6475, the low-S/N determination results in a higher metallicity than the high-S/N determination, although the values agree within two standard deviations.
For NGC~188, NGC~2243, and NGC~6791, the one or two low-S/N metallicities are lower than the one or two high-S/N ones for each cluster, but they agree within one standard deviation.
For Berkeley~29 and NGC~2682, the one or two low-S/N determinations available for each cluster lie within the range of the high-S/N mean metallicities.
Thus, spectra with a S/N as low as 20 might be sufficient to determine reliable cluster metallicities. However, the limited number of cases for comparison does not enable us to draw a firm conclusion.

\begin{table*}
\caption{Metallicities and references for low-S/N studies for clusters discussed in Sect.~\ref{sect:mean}, last paragraph.}
\label{tab:mean2}
\centering
\begin{tabular}{lrrrrrll}
\hline\hline
\noalign{\smallskip}
Cluster ID & mean \FeH & std. dev. & resolution & S/N & \# & star type & reference \\
\noalign{\smallskip}
\hline
\noalign{\smallskip}
Berkeley~29&$-$0.32&0.03&45000&25--50&5&giant&                   \citet{2008AaA...488..943S} \\
NGC~188&$-$0.12&0.16&$\approx$30000&$\approx$45&7&dwarf+subgiant&\citet{1990AJ....100..710H} \\
NGC~188&0.01&0.08&35000--57000&20--35&5&dwarf&                   \citet{2003AaA...399..133R} \\
NGC 2243 & $-$0.54 & 0.10 & 19300/28800\tablefootmark{$\dagger$} & 30--40 & 76 & dwarf+giant & \citet{2013AaA...552A.136F} \\
NGC~2682 (M67)&$-$0.04&0.01&$\approx$30000&30--45&2&dwarf&       \citet{1991AJ....102.1070H} \\
NGC~2682 (M67)&0.04&0.01&28400&30--45&2&dwarf&                   \citet{1992ApJ...387..170F} \\
NGC~6475&0.11&0.03&40000&20--30&10&dwarf&                        \citet{1997MNRAS.292..252J} \\
NGC~6791&0.30&0.08&45000&40&2&turn-off&                          \citet{2009AJ....137.4949B} \\
\noalign{\smallskip}
\hline\hline
\end{tabular}
\tablefoottext{$\dagger$}{Higher $R$ between 638 and 663~nm, lower $R$ between 660 and 696~nm.}
\end{table*}

\section{Final high-resolution sample}
\label{sect:final}

To construct the final list of reference spectroscopic cluster metallicities, we retained the restrictions in spectral resolution and S/N, even thoug lower-quality spectra may also provide reliable results. For the high-quality sample we compiled the metallicity values for individual stars from the complete literature, while the lower-quality sample compiled by us is incomplete and comprises only average cluster metallicities.

Following the considerations in Sect.~\ref{sect:assessment}, we decided to restrict the temperature and gravity range of the determinations used to compute the average metallicity of each OC. We adopted the \Teff\ range 4400--6500~K with \logg\ $\geq$ 2.0 to eliminate any rapidly rotating hot dwarfs or stars with chemical peculiarities, and bright giants possibly affected by non-LTE effects.
When two or more determinations were available for the same star, with parameters on both sides of the limiting value, the star was included, but only determinations that met the constraints were included in the average metallicity (four dwarfs and ten giants in 12 clusters).
For dwarfs, the lower \Teff\ limit eliminates three stars, without any critical consequence on the corresponding OCs (the metallicity of these cool dwarfs is very uncertain anyway).
The \Teff\ restriction is important to keep in mind for future metallicity calibrations, that is, the colour interval where they are valid.
The \Teff\ restriction removed several clusters from the sample, namely Collinder~121 and NGC~2141 (see Sect.~\ref{sect:other}), as well as Pfleiderer~4, Trumpler~2, Trumpler~20, NGC~1883, NGC~2158, and the old OC NGC~2243. The latter is supposedly one of the most metal-deficient OCs according to the value of \FeH=$-$0.48~dex determined by \mbox{\citet{1994AaA...283..911G}} for two bright giants. 

Our final list includes 458 stars in 78 OCs, with 641 metallicity determinations corresponding to 86 papers.
The measurements that are included in the final high-resolution sample are identified in Table~\ref{tab:appendix_members}.
We computed the average metallicity from all determinations for each cluster, weighted by the inverse square of the individual errors quoted by the authors.
For a significant number of determinations the authors did not quote any uncertainties. In these cases, we assumed an error of 0.1~dex.

This approach might seem problematic because authors did not compute the errors in the same way -- some quoted only internal errors while others took some external sources of uncertainty into account.
Therefore we tested several additional approaches to compute the average metallicity: a) setting the lower limit of the errors used for weighting to 0.1~dex (that is, the typical uncertainty due to the uncertainties in stellar parameters, see Sect.~\ref{sect:errors}), and b) using equal weights for all metallicity values.
The average metallicity values do not change by more than 0.03~dex for both approaches (a and b) for most clusters. They change by less than 0.1~dex for five and eight clusters (approaches a and b, respectively), with no systematic trend with metallicity. The standard deviations become larger (by more than 0.02~dex) for three clusters, and smaller for one cluster.

We tested another approach for computing the average cluster metallicity -- a two-step mean, where we first calculated the mean metallicity for each star (we recall that a significant number of stars have been analysed by several authors), and then the mean for each cluster. Using this approach, only six clusters deviate from the nominal approach (by 0.02~dex), which shows that a straight one-step average is justified.

\begin{figure*}[t]
\includegraphics[width=\textwidth,trim=0 180 0 280,clip]{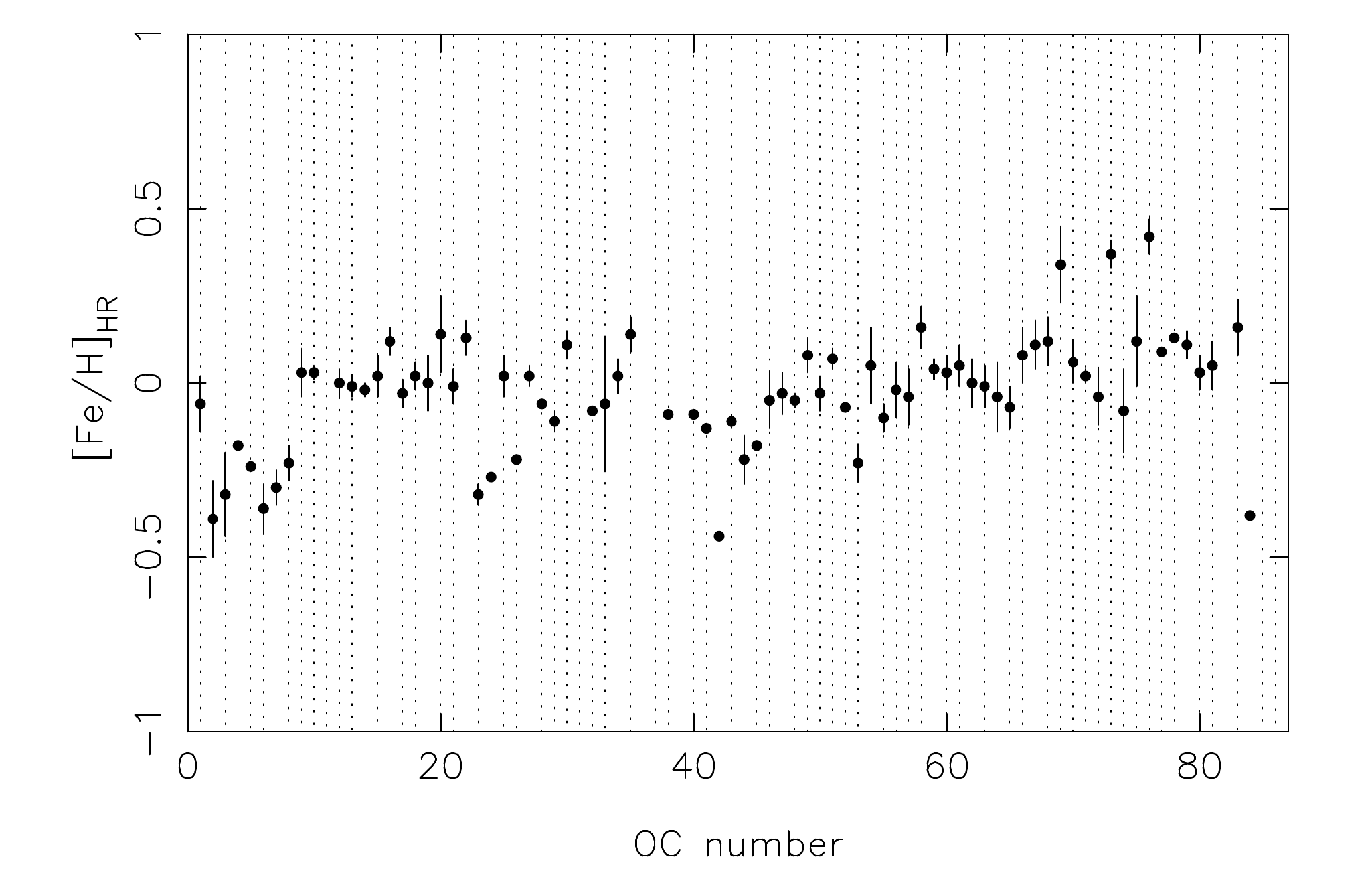}
\caption{Weighted average metallicity of OCs as listed in Table~\ref{t:mean_feh}. The error bars represent the standard deviation of the mean.}
\label{f:mean_feh}
\end{figure*}

\begin{figure}
\includegraphics[width=\columnwidth,trim=135 40 150 50,clip]{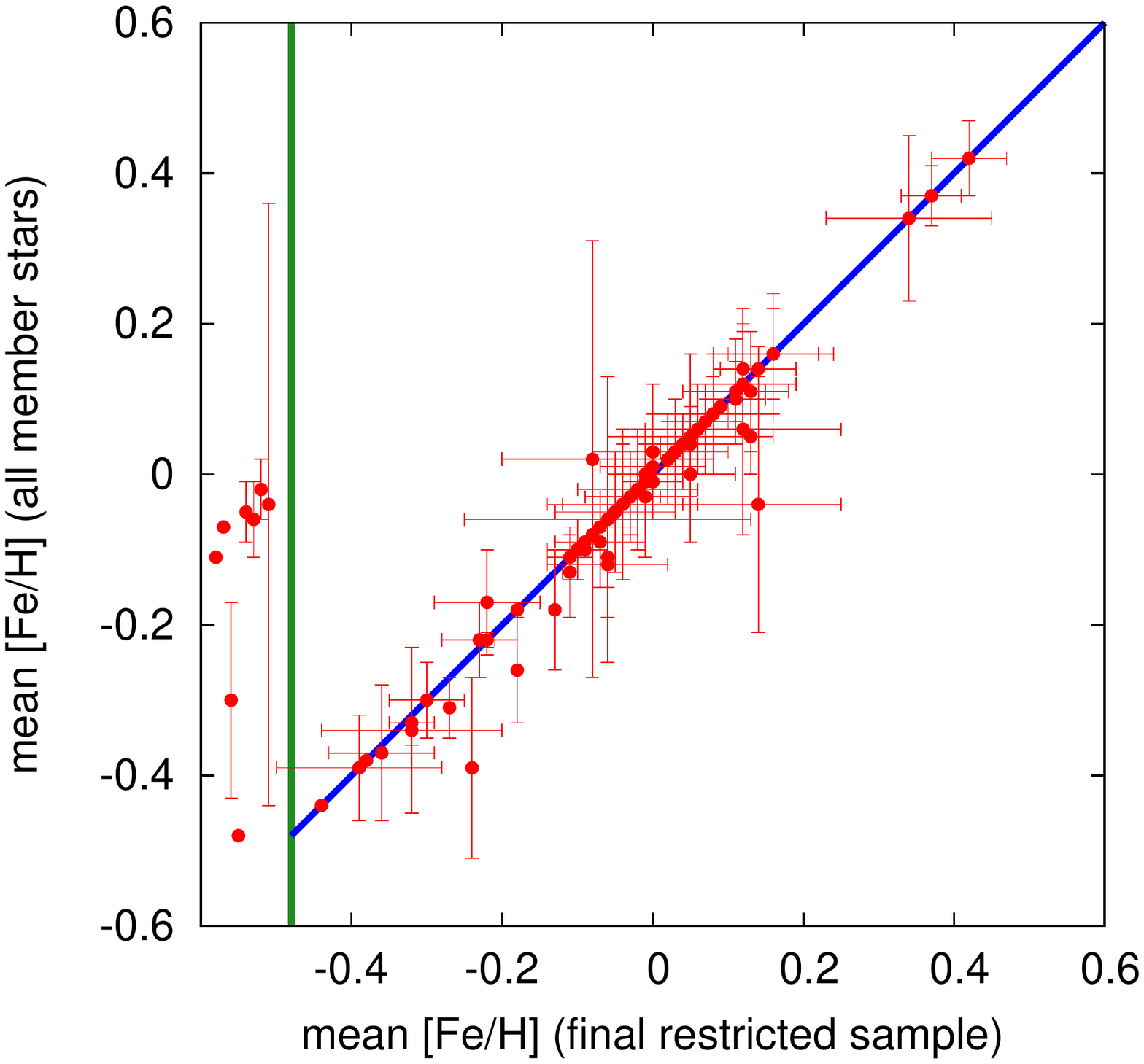}
\caption{Weighted average metallicity of OCs as listed in Table~\ref{t:mean_feh} versus weighted average metallicity of OCs, including all member stars from all publications in Table~\ref{t:mean_paper}. The error bars represent the standard deviation of the mean. Clusters to the left of the green vertical line do not appear in the final sample.}
\label{f:mean_before_after}
\end{figure}

The resulting mean metallicities (one-step average with weights using the original errors) are plotted in Fig.~\ref{f:mean_feh} and listed in Table~\ref{t:mean_feh}. The figure can be directly compared with Fig.~\ref{f:mean_paper}.
For several clusters the standard deviations shown in Fig.~\ref{f:mean_feh} are smaller than the dispersion between different determinations for one and the same cluster seen in Fig.~\ref{f:mean_paper}.
This is most evident for no. 22 -- Melotte~25 (the Hyades), and no. 74 -- NGC~6633. For the Hyades, the largest number of metallicity determinations are available (129 in Table~\ref{t:mean_paper}). After discarding 30\% probably unreliable determinations, we are still left with a significant number (92 in Table~\ref{t:mean_feh}), resulting in good statistics. 
The case of NGC~6633 is discussed in Sect.~\ref{sect:mean}.
Other clusters with improved dispersions are no. 6 -- Berkeley~29, no. 12 -- Collinder~261,  no. 16 -- IC~4651, no. 58 -- NGC~2632, no. 60 -- NGC~2682 (M67), no. 63 -- NGC~3680, and no. 80 -- NGC~752.
In Fig.~\ref{f:mean_before_after} we compare the average metallicities for the final sample with the weighted average metallicities per cluster for the starting sample, which includes all determinations for all highly probable members without restrictions on atmospheric parameters.
The mean difference (final sample minus all members) is $0.01\pm0.04$~dex, with minimum and maximum differences of $-0.10$ and +0.18~dex, respectively.
This shows that the metallicities do not change significantly by restricting the \Teff\ and \logg\ ranges, except for a few clusters with large standard deviations. The cluster changing by the largest amount (+0.18~dex) is Melotte~20 ($\alpha$ Per cluster), discussed in Sect.~\ref{sect:other}.

\begin{table*}
   \caption{Weighted average metallicity, and standard deviation, of OCs after restricting the \Teff\ and \logg\ ranges, with same running number for OCs as in Table~\ref{t:mean_paper}. See also Fig.~\ref{f:mean_feh}.}
\label{t:mean_feh}
\begin{tabular}{rllrr rllrr}
\hline\hline
\noalign{\smallskip}
OC no. & Name & Mean \FeH  &  \# &  $n$ & OC no. & Name & Mean \FeH  &  \# &  $n$ \cr
\hline
\noalign{\smallskip}
 1 & Berkeley17       &    $-$0.06 $\pm$ 0.08  &  2 &  1 &  45 & NGC2335          &    $-$0.18             &  1 &  1 \cr
 2 & Berkeley18       &    $-$0.39 $\pm$ 0.11  &  2 &  1 &  46 & NGC2355          &    $-$0.05 $\pm$ 0.08  &  3  &  3 \cr
 3 & Berkeley20       &    $-$0.32 $\pm$ 0.12  &  2 &  2 &  47 & NGC2360          &    $-$0.03 $\pm$ 0.06  & 16  &  9 \cr
 4 & Berkeley21       &    $-$0.18             &  1 &  1 &  48 & NGC2420          &    $-$0.05 $\pm$ 0.02  &  3  &  3 \cr
 5 & Berkeley22       &    $-$0.24             &  1 &  1 &  49 & NGC2423          &    $+$0.08 $\pm$ 0.05  &  6  &  3 \cr
 6 & Berkeley29       &    $-$0.36 $\pm$ 0.07  &  5 &  4 &  50 & NGC2447          &    $-$0.03 $\pm$ 0.05  & 18  &  3 \cr
 7 & Berkeley32       &    $-$0.30 $\pm$ 0.06  & 15 & 11 &  51 & NGC2477          &    $+$0.07 $\pm$ 0.03  &  4  &  4 \cr
 8 & Berkeley39       &    $-$0.23 $\pm$ 0.05  &  9 &  7 &  52 & NGC2482          &    $-$0.07             &  1 &  1 \cr
 9 & Blanco1          &    $+$0.03 $\pm$ 0.07  &  6 &  6 &  53 & NGC2506          &    $-$0.23 $\pm$ 0.05  &  7  &  5 \cr
10 & Collinder110     &    $+$0.03 $\pm$ 0.02  &  3 &  3 &  54 & NGC2516          &    $+$0.05 $\pm$ 0.11  &  2  &  2 \cr
12 & Collinder261     &    $+$0.00 $\pm$ 0.04  & 14 & 13 &  55 & NGC2527          &    $-$0.10 $\pm$ 0.04  &  2  &  2 \cr
13 & IC2391           &    $-$0.01 $\pm$ 0.03  & 14 & 12 &  56 & NGC2539          &    $-$0.02 $\pm$ 0.08  &  8  &  4 \cr
14 & IC2602           &    $-$0.02 $\pm$ 0.02  & 10 &  7 &  57 & NGC2567          &    $-$0.04 $\pm$ 0.08  &  6  &  3 \cr
15 & IC2714           &    $+$0.02 $\pm$ 0.06  &  7 &  4 &  58 & NGC2632          &    $+$0.20 $\pm$ 0.09  & 14  & 14 \cr
16 & IC4651           &    $+$0.12 $\pm$ 0.04  & 35 & 18 &  59 & NGC2660          &    $+$0.04 $\pm$ 0.03  &  4  &  4 \cr
17 & IC4665           &    $-$0.03 $\pm$ 0.04  & 18 & 18 &  60 & NGC2682          &    $+$0.00 $\pm$ 0.06  & 52  & 27 \cr
18 & IC4756           &    $+$0.02 $\pm$ 0.04  & 22 & 15 &  61 & NGC3114          &    $+$0.05 $\pm$ 0.06  &  3  &  2 \cr
19 & Melotte111       &    $+$0.00 $\pm$ 0.08  & 13 & 10 &  62 & NGC3532          &    $+$0.00 $\pm$ 0.07  & 10  &  4 \cr
20 & Melotte20        &    $+$0.14 $\pm$ 0.11  &  2 &  2 &  63 & NGC3680          &    $-$0.01 $\pm$ 0.06  & 19  & 10 \cr
21 & Melotte22        &    $-$0.01 $\pm$ 0.05  & 12 & 10 &  64 & NGC3960          &    $-$0.04 $\pm$ 0.10  &  5  &  5 \cr
22 & Melotte25        &    $+$0.13 $\pm$ 0.06  & 92 & 61 &  65 & NGC4349          &    $-$0.07 $\pm$ 0.06  &  3  &  2 \cr
23 & Melotte66        &    $-$0.32 $\pm$ 0.03  &  6 &  6 &  66 & NGC5822          &    $+$0.08 $\pm$ 0.08  & 10  &  7 \cr
24 & Melotte71        &    $-$0.27             &  1 &  1 &  67 & NGC6134          &    $+$0.11 $\pm$ 0.07  &  8  &  8 \cr
25 & NGC1039          &    $+$0.02 $\pm$ 0.06  &  7 &  7 &  68 & NGC6192          &    $+$0.12 $\pm$ 0.07  &  3  &  3 \cr
26 & NGC1193          &    $-$0.22 $\pm$ 0.01  &  2 &  1 &  69 & NGC6253          &    $+$0.34 $\pm$ 0.11  & 12  & 10 \cr
27 & NGC1245          &    $+$0.02 $\pm$ 0.03  &  3 &  3 &  70 & NGC6281          &    $+$0.06 $\pm$ 0.06  &  2  &  2 \cr
28 & NGC1545          &    $-$0.06             &  1 &  1 &  71 & NGC6475          &    $+$0.02 $\pm$ 0.02  &  3  &  3 \cr
29 & NGC1817          &    $-$0.11 $\pm$ 0.03  &  7 &  4 &  72 & NGC6494          &    $-$0.04 $\pm$ 0.08  &  6  &  3 \cr
30 & NGC188           &    $+$0.11 $\pm$ 0.04  &  8 &  4 &  73 & NGC6583          &    $+$0.37 $\pm$ 0.04  &  2  &  2 \cr
32 & NGC1901          &    $-$0.08             &  1 &  1 &  74 & NGC6633          &    $-$0.08 $\pm$ 0.12  &  8  &  4 \cr
33 & NGC1977          &    $-$0.06 $\pm$ 0.19  &  2 &  2 &  75 & NGC6705          &    $+$0.12 $\pm$ 0.13  & 10  &  7 \cr
34 & NGC2099          &    $+$0.02 $\pm$ 0.05  &  3 &  3 &  76 & NGC6791          &    $+$0.42 $\pm$ 0.05  &  8  &  8 \cr
35 & NGC2112          &    $+$0.14 $\pm$ 0.05  &  3 &  3 &  77 & NGC6819          &    $+$0.09 $\pm$ 0.01  &  3  &  3 \cr
38 & NGC2194          &    $-$0.09 $\pm$ 0.00  &  2 &  2 &  78 & NGC6939          &    $+$0.13             &  1 &  1 \cr
40 & NGC2251          &    $-$0.09             &  1 &  1 &  79 & NGC7142          &    $+$0.11 $\pm$ 0.04  &  7  &  4 \cr
41 & NGC2264          &    $-$0.13             &  1 &  1 &  80 & NGC752           &    $-$0.02 $\pm$ 0.04  & 19  & 18 \cr
42 & NGC2266          &    $-$0.44             &  1 &  1 &  81 & NGC7789          &    $+$0.01 $\pm$ 0.04  &  6  &  5 \cr
43 & NGC2287          &    $-$0.11 $\pm$ 0.01  &  3 &  2 &  83 & Ruprecht147      &    $+$0.16 $\pm$ 0.08  &  5  &  5 \cr
44 & NGC2324          &    $-$0.22 $\pm$ 0.07  &  2 &  2 &  84 & Saurer1          &    $-$0.38 $\pm$ 0.00  &  2  &  2 \cr

\noalign{\smallskip}
\hline\hline
\noalign{\smallskip}
\end{tabular}
\tablefoot{
The column headed ``\#'' gives the number of metallicity determinations, and the column headed ``$n$'' the number of individual member stars.
}
\end{table*}

At the metal-rich end of the metallicity distribution of the final sample, we find three OCs with extreme values: NGC~6253 (\FeH$\approx+0.3$~dex; \mbox{\citealt{2007AaA...473..129C}}; \mbox{\citealt{2007AaA...465..185S}}; \mbox{\citealt{2012MNRAS.423.3039M}}), NGC~6583 \mbox{\citep[\FeH$\approx+0.4$~dex;][]{2010AaA...523A..11M}}, and NGC~6791 \citep[\FeH$\approx+0.4$~dex;][]{2006ApJ...642..462G,2012ApJ...756L..40G}.
At the metal-poor end, there are eight OCs with metallicities of about $-$0.3~dex and below: Berkeley~18, Berkeley~20, Berkeley~29, Berkeley~32, Melotte~66, Melotte~71, NGC~2266, and Saurer~1.
The metallicities of three of these clusters rely on a substantial number of stars:
11 stars in Berkeley~32, leading to \FeH=$-$0.30 $\pm$0.06 (\mbox{\citealt{2006AaA...458..121S}}; \mbox{\citealt{2011AaA...535A..30C}}; \mbox{\citealt{2012AJ....144...95Y}}; \mbox{\citealt{2013AJ....145..107J}}), six stars in Melotte~66, leading to \FeH=$-$0.32 $\pm$0.03 \mbox{\citep{2008AaA...488..943S}}, and four stars in Berkeley~29, leading to \FeH=$-$0.36 $\pm$0.07 \mbox{\citep{2004AJ....128.1676C,2008AaA...488..943S}}.
On the contrary, the metallicity of Melotte~71 (\FeH=$-$0.27) is based on only one very faint star, PJ~127, $V\simeq17$, which has \Teff=4610~K, \logg=2.16, and an uncertainty on the metallicity of 0.24 \mbox{\citep{1996AJ....112.1551B}}.
NGC~2266 is the cluster with the lowest metallicity, \FeH=$-$0.44, which is also based on only one star, although a brighter one ($V\simeq11$) and with a lower uncertainty on the metallicity of 0.05 \mbox{\citep{2013MNRAS.431.3338R}}. 
It is worth mentioning that the authors argued that this cluster may be part of the thick disk, based on its space motions.
The low metallicity of Berkeley~18 is also based on one faint star, but two measurements (star K 1163 with $V\simeq16$, see Sect.~\ref{sect:metalpoor}).
The remaining two of the most-metal-poor clusters (Berkeley~20 and Saurer~1) have metallicities relying on two stars.
We have eleven more OCs in the final sample with metallicities relying on only one star, most of them with subsolar metallicities.

For the three clusters with the largest number of determinations ($>30$), we can separate the stars into giants (\logg $\le$ 3.0) and dwarfs (cf. Sect.~\ref{sect:m67} for M67).
The weighted mean metallicities for the separated samples and the number of determinations are listed in Table~\ref{tab:dwarfsgiants}.
The determinations for the Hyades (Melotte~25) are mainly for dwarfs. For M67 (NGC~2682) and IC~4651, the determination numbers are more similar for giants and dwarfs.
We found no indication for a metallicity difference between the dwarfs and giants for any of these clusters.

\begin{figure}
\includegraphics[width=\columnwidth,trim=135 40 150 50,clip]{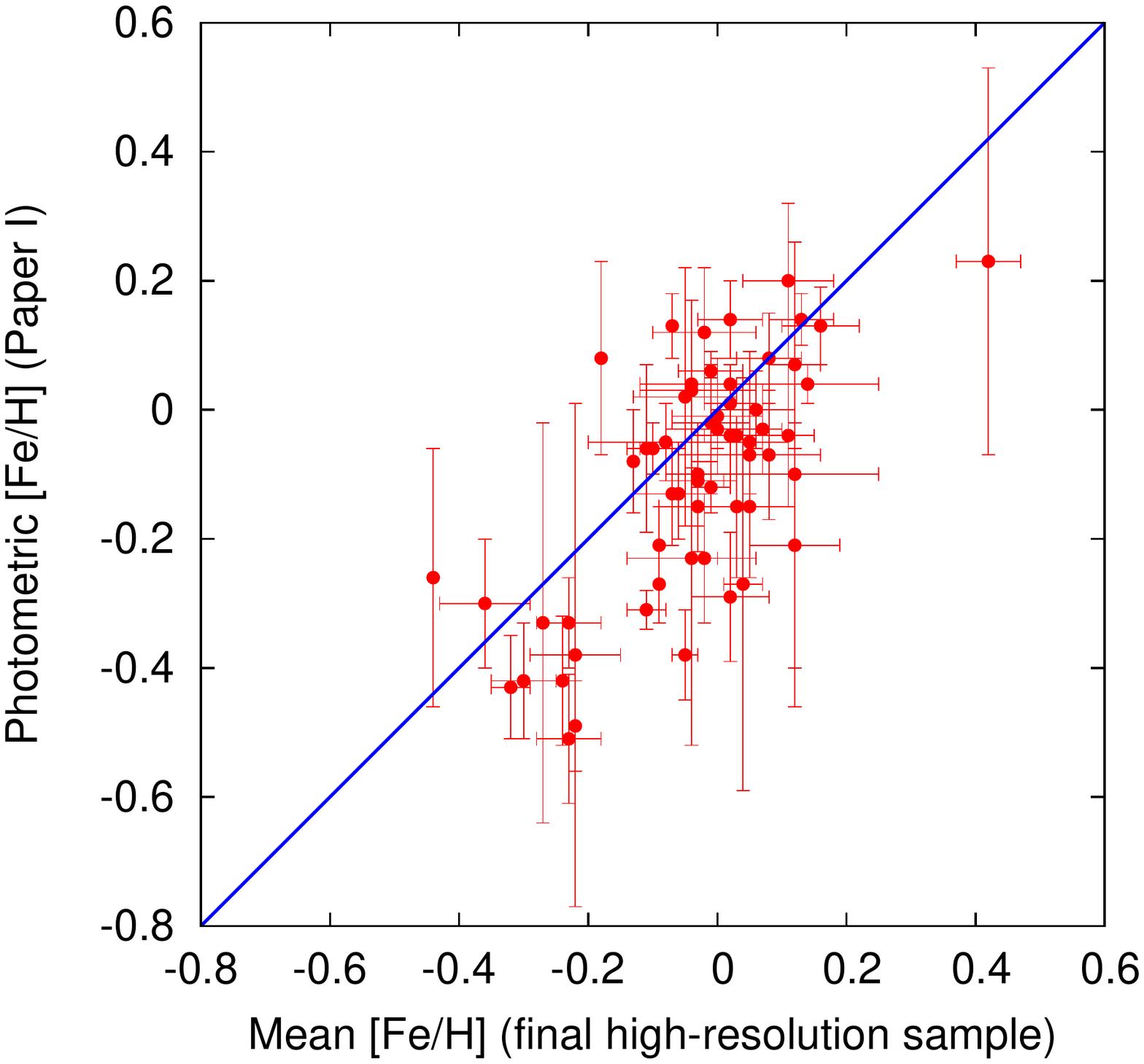}
\caption{Comparison of mean metallicities listed in Table~\ref{t:mean_feh} and those obtained from photometry in Paper~I for 62 OCs in common (two clusters lie outside the axis ranges -- no. 35, NGC~2112, and no. 4, Berkeley~21).}
\label{f:HR-phot}
\end{figure}

We compare in Fig.~\ref{f:HR-phot} the mean metallicities determined from high-resolution spectroscopy and that obtained from photometry (Paper~I).
There are 62 clusters in common, and the mean difference between photometric and spectroscopic values is $-0.10$~dex with a median of $-0.08$ and a standard deviation of 0.23.
The two most extreme cases are NGC~2112 and Berkeley~21, with differences of $-1.44$ and $-0.78$~dex, respectively. These are also two of the most metal-poor clusters in the photometric sample ([Fe/H]$_{\rm phot}= -1.30$ and $-0.96$~dex, respectively).
There are no other clusters with photometric metallicities below $-0.5$~dex in the common sample.
We conclude that the photometric metallicity scale is more metal-poor than the spectroscopic one.

Combining our two samples of clusters with photometric and spectroscopic metallicities would result in a total number of 204 clusters with known metallicity. However, the differences in metallicity found for the two samples and the large intrinsic dispersion of the photometric determinations requires a calibration of the photometric sample to the spectroscopic metallicity scale. This process and the discussion of the full sample will be presented in a forthcoming article.

\begin{table}
   \caption{Weighted mean metallicities and standard deviations for giant (\logg $\le$ 3.0) and dwarf samples in three clusters.}
\label{tab:dwarfsgiants}
\centering
\begin{tabular}{lrr}
\hline\hline
\noalign{\smallskip}
Name & Mean \FeH & \#  \\
\noalign{\smallskip}
\hline
\noalign{\smallskip}
Melotte~25 giants    & 0.12 $\pm$ 0.04 & 16 \\
Melotte~25 dwarfs    & 0.13 $\pm$ 0.06 & 76 \\ 
\noalign{\smallskip}
\hline
\noalign{\smallskip}
NGC~2682 giants      & $-$0.05 $\pm$ 0.05 & 23 \\
NGC~2682 dwarfs      &    0.01 $\pm$ 0.06 & 29 \\
\noalign{\smallskip}
\hline
\noalign{\smallskip}
IC~4651 giants       & 0.12 $\pm$ 0.05 & 15 \\
IC~4651 dwarfs       & 0.12 $\pm$ 0.04 & 20 \\
\noalign{\smallskip}
\hline\hline
\end{tabular}
\end{table}

\section{Discussion}
\label{sect:discussion}
\subsection{Comparison with other metallicity studies}
\label{sect:discusscompare}


\begin{figure}
\includegraphics[width=\columnwidth,trim=135 40 150 50,clip]{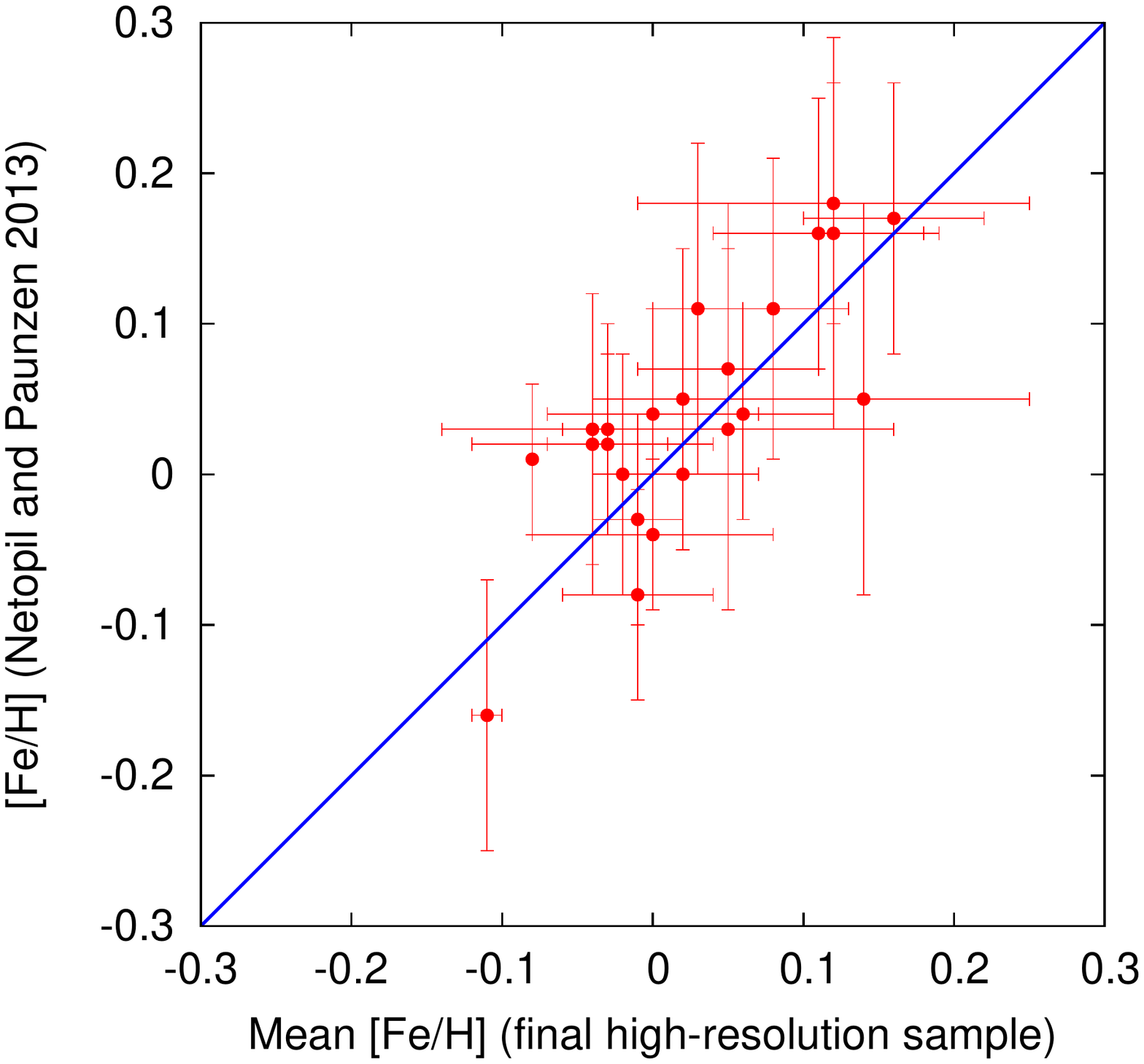}
\caption{Weighted average metallicity of OCs as listed in Table~\ref{t:mean_feh} versus metallicities determined by \citet{2013arXiv1307.2094N}.}
\label{f:final_NP13}
\end{figure}

\mbox{\citet{2013arXiv1307.2094N}} determined metallicities of 58 OCs from photometry, using an isochrone-fitting method developed by \mbox{\citet{2010AaA...514A..81P}}.
In Fig.~\ref{f:final_NP13} we compare the mean metallicities of the final high-resolution sample (Table~\ref{t:mean_feh}) with the metallicities of \mbox{\citet{2013arXiv1307.2094N}}.
We have 23 clusters in common with their sample, all of which have metallicities between $-0.1$ and +0.2~dex.
The mean difference between the photometric and spectroscopic values is $-0.02\pm~0.05$~dex, with a maximum absolute difference of 0.09~dex. 
We confirm the conclusion of \mbox{\citet{2013arXiv1307.2094N}} that their tool is useful for estimating OC metallicities from photometry.


\begin{figure}
\includegraphics[width=\columnwidth,trim=50 40 60 60,clip]{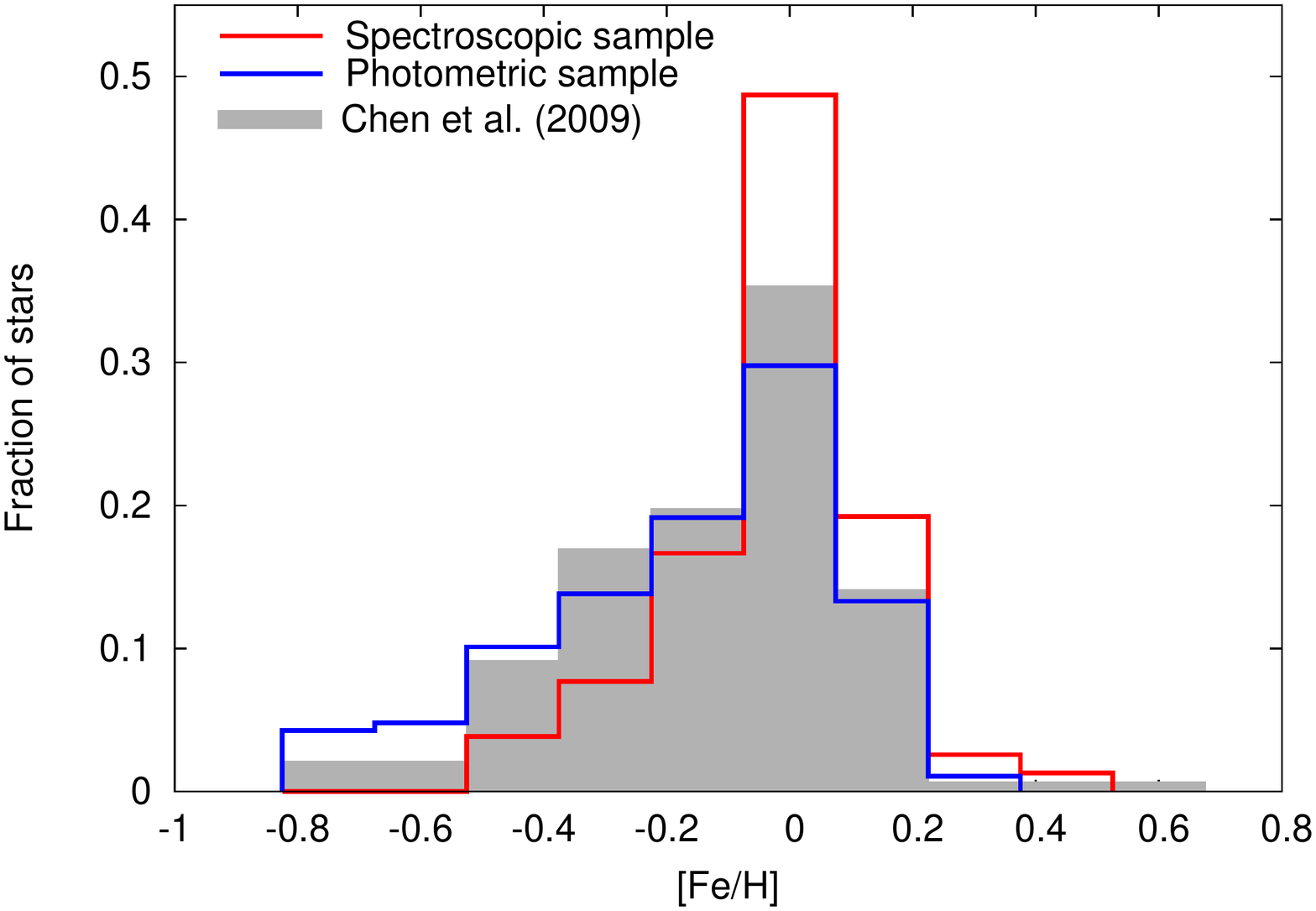}
\caption{Distribution of metallicities for the final spectroscopic sample (red line), the photometric sample (Paper I, blue line), and the sample of \citet[grey full histogram]{2009IAUS..254P..15C}.}
\label{f:methistogram}
\end{figure}

A recent compilation of data, including metallicities, for a large number of OCs is the catalogue of \mbox{\citet{2003AJ....125.1397C}}. A preliminary update is presented in \mbox{\citet{2009IAUS..254P..15C}}, which contains 144 clusters with metallicity, distance, and age values. However, the metallicities are a mixture of photometric and spectroscopic determinations.
Fig.~\ref{f:methistogram} shows the histogram of cluster metallicities from \mbox{\citet{2009IAUS..254P..15C}} together with those of our spectroscopic and photometric samples. The figure indicates that the peak at solar metallicity in the histogram of \mbox{\citet{2009IAUS..254P..15C}} may be dominated by spectroscopic metallicities, while the metal-poor tail is mainly due to photometric metallicities.


\begin{figure*}
\includegraphics[width=\textwidth,trim=0 20 0 0,clip]{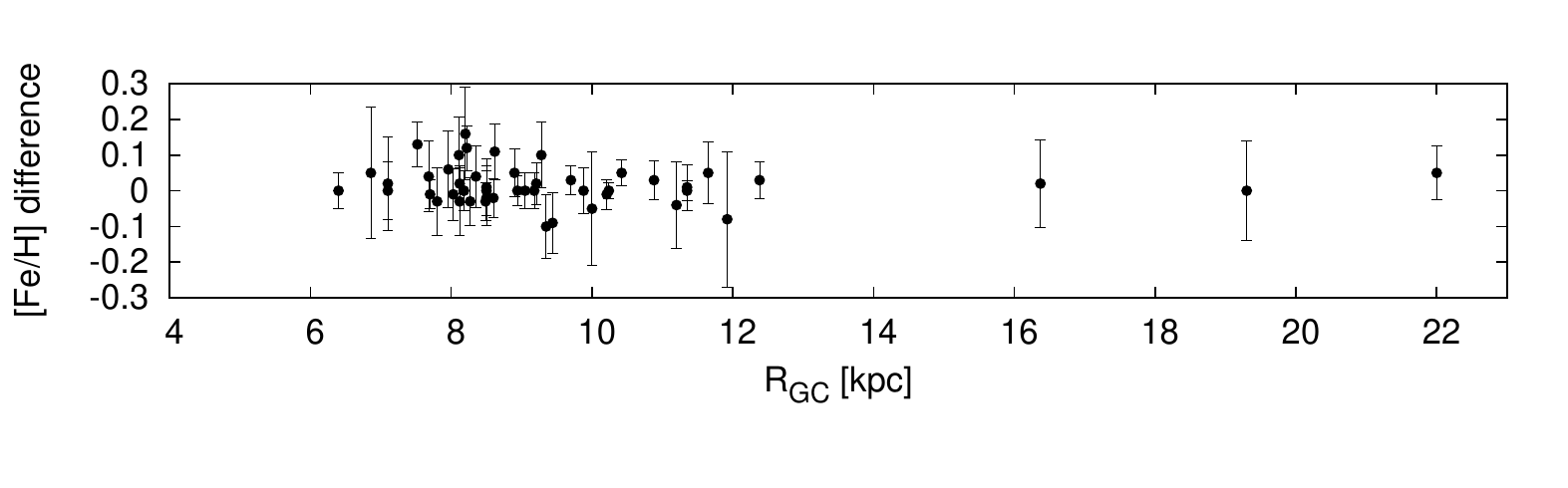}
\caption{Comparison of mean metallicities listed in Table~\ref{t:mean_feh} and those compiled by \citet{2010AaA...523A..11M}, for 49 OCs in common, as a function of Galactocentric distance. The points represent [Fe/H](Magrini) minus [Fe/H](this work). The error bars represent the standard deviations of the cluster means from both lists, added in quadrature. This can be directly compared with Fig.~9 of \citet{2010AaA...523A..11M}.}
\label{f:MagriniHeiterCommon}
\end{figure*}

In Fig.~\ref{f:MagriniHeiterCommon} we compare the mean metallicities of our final high-resolution sample with those compiled by \mbox{\citet[][see Sect.~\ref{sect:introduction}]{2010AaA...523A..11M}}.
The mean difference for all clusters in common is 0.02~dex (median 0.00~dex), with a standard deviation of 0.05~dex. However, the dispersion varies with $R_{\rm GC}$, and the lowest and highest values of the deviations are $-0.10$ and +0.16~dex, respectively.
The scatter seen in this figure should be added to Fig.~9 of \mbox{\citet{2010AaA...523A..11M}}, representing the uncertainty originating from the way of combining cluster abundances from different authors. For example, the scatter is 0.06~dex for $7.5 \lesssim R_{\rm GC}\lesssim 9.5$~kpc, adding to the uncertainty of the inner disk gradient determined by \mbox{\citet{2010AaA...523A..11M}}.

We have 64 clusters in common with the sample of \mbox{\citet[][see Sect.~\ref{sect:introduction}]{2011AaA...535A..30C}}. 
We do not show a detailed comparison of metallicities, since their sample is very similar to ours, and we included all of their references that meet our constraints.
The differences are that their metallicities are based on published cluster means, not on individual stars, and that they set a lower limit on $R$ (15000) than we do, and no limit on S/N (spectra with S/N as low as 5 are included).


Ultimately, the OC metallicities should be combined with other tracers of metallicity in the Galaxy, such as HII regions, B-type stars, planetary nebulae, and Cepheids. However, such combined samples will have to deal with differences in metallicity scale, not only for the same type of objects due to different methods, but also between different types of objects.
A large sample of Cepheids has been used by \mbox{\citet{2009AaA...504...81P}} to study the metallicity gradient of the Galactic disk. They compared the metallicities of their sample as a function of Galactocentric distance with several samples of OCs. They found significant differences in derived metallicity gradients, which were partly attributed to different Galactocentric distributions of these tracers.

It is worth noting that \mbox{\citet{2009AaA...504...81P}} combined spectroscopic metallicities for over 200 Cepheids with photometric ones for about 60. The mean difference between photometric and spectroscopic metallicities ($-0.03$~dex) and the intrinsic dispersion (0.15~dex) are smaller than for our two samples. This could partly be due to the more homogeneous photometric metallicities determined by \mbox{\citet{2009AaA...504...81P}} from one set of photometric bands (Walraven and K-band) and a metallicity calibration based on recent stellar evolution models. However, a systematic difference between two of their sources for spectroscopic metallicities is apparent in their Fig. 3.

\subsection{Application to Galactic structure}
\label{sect:galactic}

\begin{figure}
\includegraphics[width=\columnwidth,trim=135 35 180 50,clip]{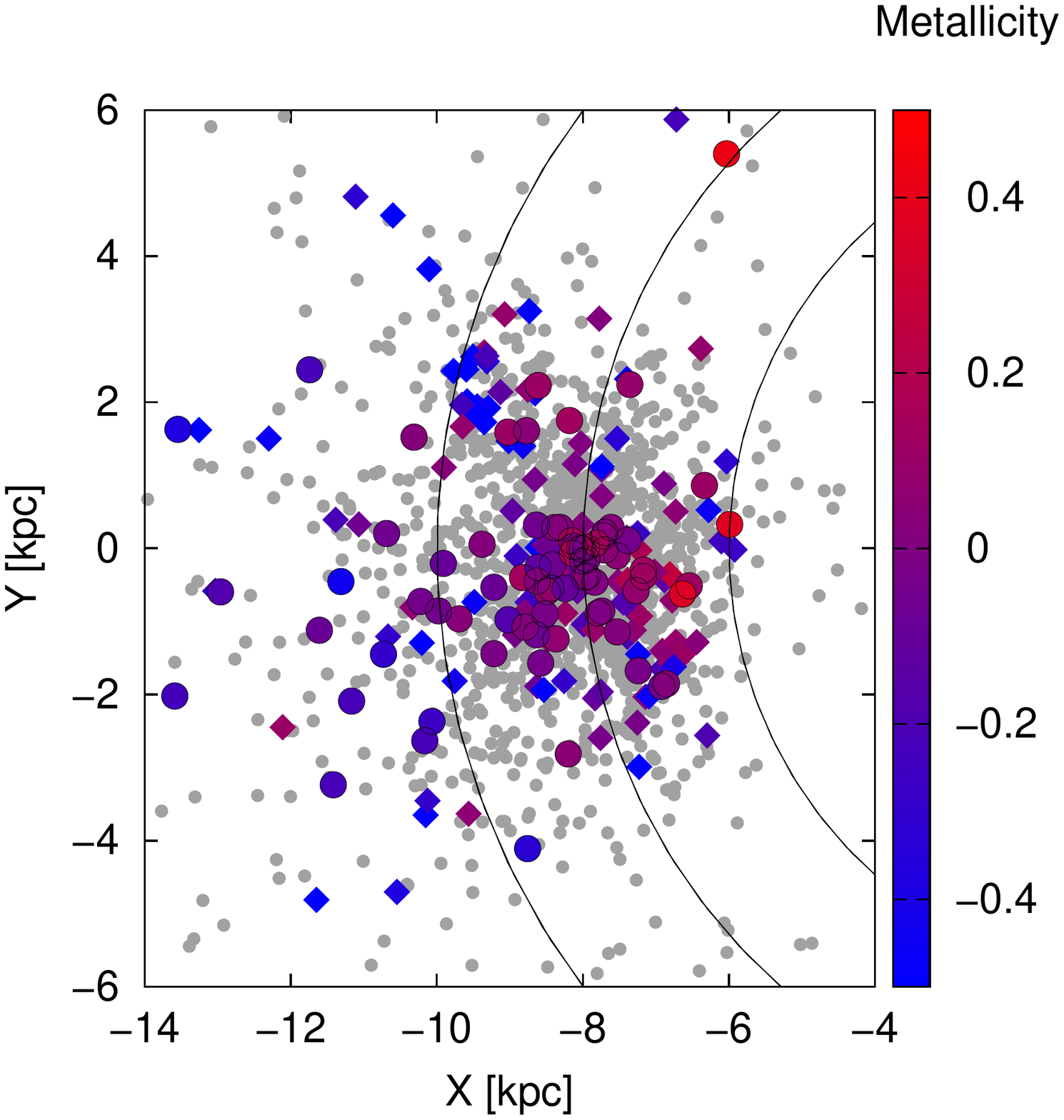}
\caption{Distributions of the OCs from Paper~I (diamonds), the OCs in the spectroscopic sample (coloured circles), and the OCs in the \citet[][version 3.2]{2002A&A...389..871D} catalogue (grey circles) projected onto the Galactic plane, in a linear coordinate system where $X$ increases from the Sun towards the Galactic centre, and $Y$ increases in the direction of Galactic rotation at the location of the Sun. The Galactic centre is located at ($X,Y$) = (0,0). Metallicity is represented by the colour scale shown in the bar to the right. Solid lines indicate distances from the Galactic centre of 6, 8, and 10~kpc. Three clusters in the spectroscopic sample and one in the photometric sample lie outside the shown $X$ range.}
\label{f:xydistribution}
\end{figure}
%


In this section we use our final sample of OCs with spectroscopic metallicities for a preliminary investigation of the distribution of metals in the Galactic disk.

In Fig.~\ref{f:xydistribution} we show the spatial distribution of OCs projected onto the Galactic plane. Together with our final spectroscopic sample, we also include the photometric sample from Paper~I and all clusters in the \mbox{\citet[][Version 3.2]{2002A&A...389..871D}} catalogue. The metallicity of each cluster in the first two samples is indicated by the colour of the symbol. We recall that the photometric and spectroscopic metallicities are not yet on the same scale. Even though a combined sample would trace the metallicity throughout the disk better than the spectroscopic one alone, gaps are evident, which should be filled by targeted observations of known clusters.
We did not attempt to derive a more detailed metallicity distribution for the spectroscopic sample in the immediate solar neighbourhood ($\pm2$~kpc), as we did in Fig.~4 of Paper~I.
The number of clusters in that area is simply too small (55 compared with 128 in Paper~I).
However, we speculate that the distribution based on a combined calibrated sample will be more smooth than that based on the photometric sample alone. In particular, the most distinct feature of Fig.~4 in Paper~I, a dip in metallicity at $X=-8.4$ and $Y=-0.4$~kpc, will disappear. It is caused by the cluster NGC~2112, which has a spectroscopic metallicity of $+0.14\pm0.08$~dex, based on three stars in two publications, while its photometric metallicity of $-1.3\pm0.2$~dex is based on Str\"omgren colours of four stars in one publication.

\begin{figure*}
\includegraphics[width=\textwidth]{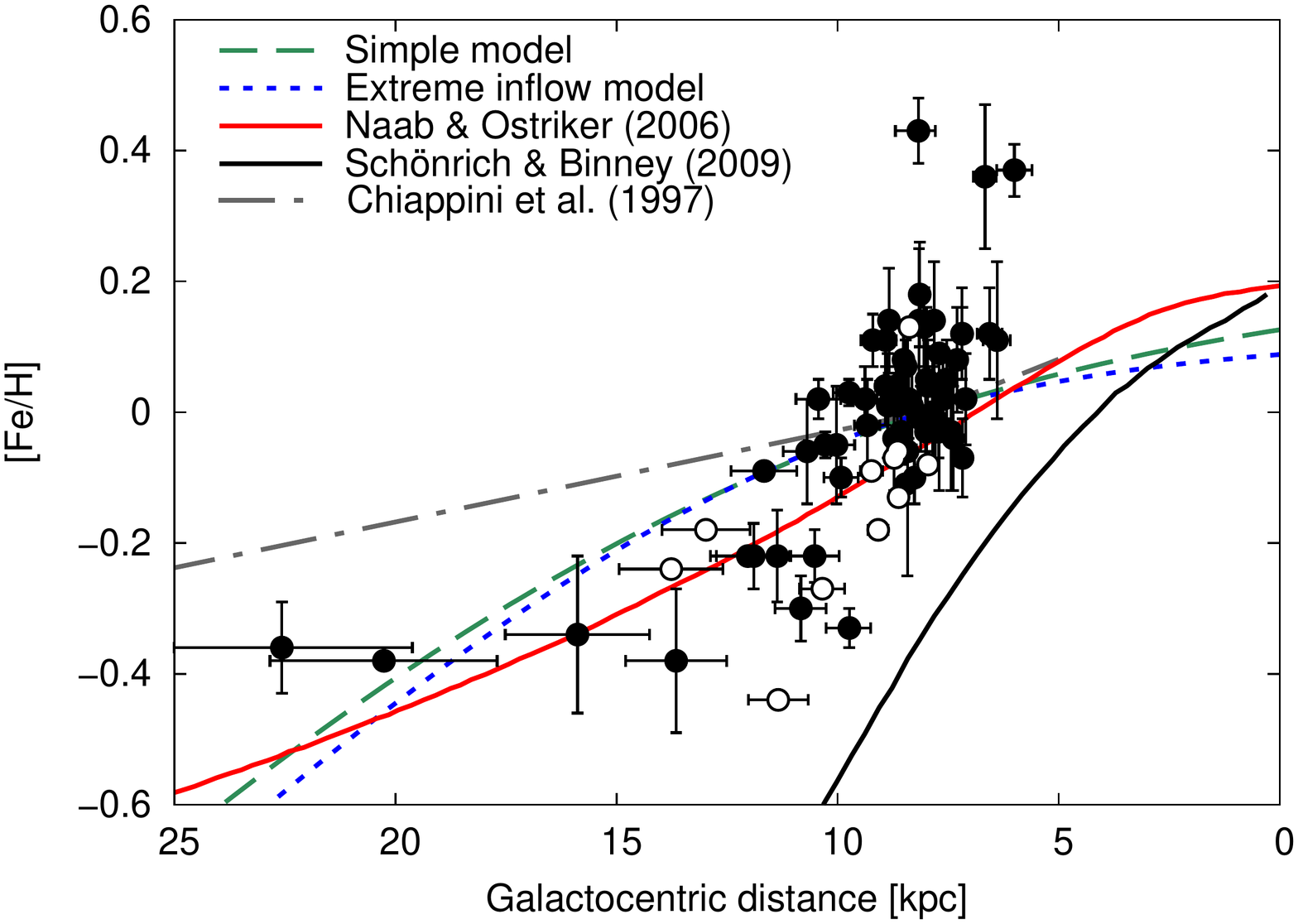}
\caption{Cluster metallicity as a function of Galactocentric distance computed from the coordinates of the clusters and their distance from the Sun taken from the \citet[][version 3.2]{2002A&A...389..871D} catalogue. The distance of the Sun from the Galactic centre (at the right of the graph) is assumed to be 8~kpc. Open symbols are clusters with only one metallicity measurement. Also shown are predictions of several models for the Galactic chemical evolution. See Sect.~\ref{sect:galactic} for details.}
\label{f:fehrgc}
\end{figure*}

Figure~\ref{f:fehrgc} shows the cluster metallicity from Table~\ref{t:mean_feh} as a function of Galactocentric distance $R_{\rm GC}$ compared with the predictions of several models for Galactic chemical evolution.
The Galactocentric distances were computed from the coordinates of the clusters and their distance from the Sun as given in the catalogue of \mbox{\citet[][version 3.2]{2002A&A...389..871D}}. The distance of the Sun from the Galactic centre is assumed to be $R_{\rm GC,\odot}$=8~kpc.
The error bars for $R_{\rm GC}$ correspond to a variation of the cluster distances by $\pm$20\%.
Most clusters are located in the range $6 \lesssim R_{\rm GC,\odot} \lesssim 10$~kpc. The clusters within this range are also mostly located close to the Sun in the Galactic plane, as can be seen in Fig.~\ref{f:xydistribution}, guided by the solid lines.

The metallicity distributions predicted by the following models are shown in Fig.~\ref{f:fehrgc}:
\begin{itemize}
   \item The \emph{simple model} with instantaneous recycling described, for example, in \mbox{\citet[][eq.~8.7]{2009nceg.book.....P}}, green long-dashed line. This model allows one to derive an analytic expression for the average abundance of Fe in a stellar population:
      \[ \langle z(Fe) \rangle = 1+\frac{\mu \ln(\mu)}{1-\mu},  \] 
      where the gas fraction $\mu = g/(s+g)$, where $g$ is the density of gas in the system, and $s$ is the density of matter in the form of stars. 
      The dependence of the gas fraction and, in turn, of the metallicity on $R_{\rm GC}$ is obtained by assuming the following relations for the gas and stellar surface densities in $M_\odot$ pc$^{-2}$ as a function of Galactocentric distance in kpc:
      \[ g = 15.0 e^{\frac{-R_{\rm GC}}{9.9}} \]
      (fit to data shown in Fig.~1 of \mbox{\citealt{1993AIPC..278..267D}}, for $R_{\rm GC} > 4.5$~kpc), and
      \[ s = 198 e^{\frac{-R_{\rm GC}}{4.0}}.  \]
      The latter was normalized to give a gas fraction of 0.2 at the distance of the Sun (see \mbox{\citealt{2009nceg.book.....P}}, Table~7.9). 
   \item The \emph{extreme inflow model}, with a metallicity dependence derived by \mbox{\citet{1972NPhS..236....7L}} as:
      \[ \langle z(Fe) \rangle = 1+\frac{1}{s/g (e^{-s/g}-1)} \] 
      (see \citealt{2009nceg.book.....P}, eq.~8.28), blue short-dashed line.
   \item The model by \mbox{\citet{1997ApJ...477..765C}} -- gradient of iron at 12~Gyr from their Table~4, but using $R_{\rm GC,\odot}$=8~kpc, grey dash-dotted line.
   \item The model by \mbox{\citet{2006MNRAS.366..899N}} -- present-day metallicities of stars from their Fig.~11, red solid line.
   \item The model by \mbox{\citet{2009MNRAS.396..203S}} -- mean metallicities of stars at the present time from their Fig.~11, black solid line.
\end{itemize}



Even though the first two analytical models can be regarded as two opposite extreme cases, their predicted radial dependence of metallicity is quite similar. It does not agree with the observed metallicity dependence, in particular at distances between about 10 and 15~kpc.

The main feature of the two-infall model by \mbox{\citet{1997ApJ...477..765C}} is that the thick and thin disks form by accretion of extragalactic material on very different timescales (1~Gyr and 8~Gyr, respectively). The model predictions agree well with several observed properties of the Galaxy, such as the stellar metallicity distribution and the fraction of metal-poor stars in the solar neighbourhood.
However, the metallicity gradient predicted by the model is much shallower than that suggested by the OC metallicities (Fig.~\ref{f:fehrgc}).

In the model of \mbox{\citet{2006MNRAS.366..899N}}, the evolution of the gas infall rate is prescribed based on spherical infall theory and the current observed distribution of the total disk surface mass. For other ingredients (star formation, IMF, chemical evolution), the model uses standard prescriptions.
Among the models discussed here, the radial dependence of metallicity predicted by this model agrees best with the OC metallicities, except that its metallicity is somewhat low at the solar radius. However, \mbox{\citet{2006MNRAS.366..899N}} mentioned that their model predicts a significantly steeper metallicity gradient for past times. Including information on the cluster ages might improve the agreement. Also, changing the IMF in the model from \mbox{\citet{1955ApJ...121..161S}} 
to \mbox{\citet{2003PASP..115..763C}} 
leads to a steeper gradient.

The model of \mbox{\citet{2009MNRAS.396..203S}} introduces both radial gas flows and radial migration of stars, beyond the standard ingredients of chemical evolution models. This model predicts star counts as a function of various stellar parameters in agreement with observations, and it reproduces correlations between tangential velocity and abundance patterns. It also produces a thick disk alongside the thin disk within the Galaxy.
Regarding the predicted metallicity as a function of $R_{\rm GC}$, the model curve does not coincide with any of the observed OC metallicities.
To bring the model and observations into agreement would require a shift in metallicity of at least 0.3~dex.
However, the model gradient is very similar to the observed one around the solar radius ($7\lesssim R_{\rm GC} \lesssim 12$~kpc).
The steep gradient is mainly caused by the radial gas flow, as \mbox{\citet{2009MNRAS.396..203S}} showed by varying the model parameters. 
Their model also predicts distributions over metallicity of stars at fixed $R_{\rm GC}$ with a full-width at half maximum (FWHM) around 0.35~dex. 
This value agrees exactly with the FWHM of the metallicity distribution of the 26 clusters in our sample with $7\lesssim R_{\rm GC} \lesssim 9$~kpc.

Although the list of discussed models is not exhaustive, we can conclude that none of the current models for Galactic chemical evolution succeeds in predicting the metallicity gradient observed for the Galactic disk based on OCs. However, the significance of the comparison is limited by the inhomogeneous distribution of clusters over distance and the fact that the sample contains clusters of different ages. Moreover, we did not take into account possible radial migrations of the clusters.
We postpone a more detailed comparison to a forthcoming paper, where we will combine the photometric and spectroscopic metallicity determinations in a proper way.

\section{Summary and conclusions}
\label{sect:conclusions}

For this article, metallicities of individual stars in OCs resulting from spectroscopy at high resolution (R$>$25000) and high signal-to-noise ratio (S/N$>$50) were exhaustively gathered from the recent literature (publication year 1990 and later). Only the most probable members in 86 OCs were considered for further analysis. Some discrepancies in metallicities of individual stars and cluster means were found for a fraction of OCs studied by several authors. The sources of these differences were analysed in relation to the quoted errors, and various aspects of the spectroscopic analyses with an impact on metallicity were discussed.

Comparisons were also made with metallicities based on lower resolution spectra and spectral indices, and on lower S/N observations. We found that low-resolution estimates are in general more metal poor than higher-resolution determinations. At medium and high resolution ($R$$\ge$13000) and S/N$>$20, the temperature scale, the line list, the methodology, and the choice of the microturbulence parameter, seem to play a larger role for the reliability of the metallicities than the resolution and S/N. However, the largest contribution to the observed dispersion comes from the properties of the stars. Chemically peculiar stars and binaries must obviously be removed when averaging the metallicity determinations in an OC, but also bright giants that are possibly affected by non-LTE, and hot dwarfs that are possibly affected by rapid rotation. 

These considerations led us to build a clean sample of metallicity determinations of individual stars in a restricted temperature and gravity range, after rejecting some studies that appeared to be affected by systematic uncertainties. The numbers of different stars and metallicity determinations were reduced by 25\% and nearly 30\%, respectively, between the starting and the final sample. The final sample includes 458 stars with 641 metallicity determinations in 86 papers, which were used to compute the weighted average metallicity of 78 OCs. We found no difference in mean metallicities deduced from dwarfs and giants, based on three OCs with a significant number of determinations for both groups. 

Photometric metallicities compiled in Paper~I were found to be systematically more metal-poor by 0.11~dex than the spectroscopic ones presented here, with a standard deviation of 0.23~dex. However, recent photometric determinations by \mbox{\citet{2013arXiv1307.2094N}} 
agree much better. 
The compilation of spectroscopic metallicities by \mbox{\citet{2010AaA...523A..11M}}, 
which consists of the cluster metallicity from one selected publication per cluster, agrees on average with our spectroscopic metallicities for the clusters in common. However, metallicities for individual clusters deviate by up to 0.16~dex.  

We used our final sample to test four models that predict the radial metallicity gradient in the Galaxy. None of them was found to fully agree with the OC metallicities versus Galactocentric distance. The model by \mbox{\citet{2009MNRAS.396..203S}} 
has a similar slope, but shows a metallicity shift of 0.3~dex. The metallicity dispersion at the solar radius predicted by this model is similar to the one measured for our sample. 
This comparison shows that existing models of Galactic chemical evolution cannot reproduce the currently available sample of cluster metallicities even within the rather large observational uncertainties, and calls for further developments on the theoretical side.

This work demonstrates, however, that it is crucial to enlarge the number of OCs with accurate metallicities from spectroscopic studies. Metallicity determinations are needed in particular at small and large Galactocentric distances, and in several regions of the Galactic disk that are poorly sampled. To obtain reliable cluster metallicities will require an increase in the number of individual stars studied in each cluster, a careful selection of the stars, and a homogeneous analysis of the spectra. This is one of the main aims of the Gaia-ESO Public Spectroscopic Survey, which will dramatically improve the situation of spectroscopic metallicities of OCs, with 100 clusters to be observed in the next four years.

Studies of Galactic structure should take advantage of the large number of existing photometric metallicities for OCs (Paper~I). In combination with the current spectroscopic sample, a sample of more than 200 clusters with known metallicities can be constructed.
A calibration of the photometric sample using the spectroscopic sample will be required to define a common metallicity scale with a small intrinsic dispersion.
The details of this calibration and the impact of the combined sample will be the subject of the next paper of this series.

\begin{acknowledgements}
UH acknowledges support from the Swedish National Space Board (Rymdstyrelsen). We thank the Bordeaux 1 University for providing UH a one-month position in 2009 and 2010.
This work was supported by the SoMoPro II Programme (3SGA5916), co-financed by the European Union and the South Moravian Region, grant GA \v{C}R 7AMB12AT003, and the financial contributions of the Austrian Agency for International Cooperation in Education and Research (CZ-10/2012).
We thank the referee, Elena Pancino, for valuable comments.
This research has made use of the WEBDA database, operated at the Department of Theoretical Physics and Astrophysics of the Masaryk University, and of the SIMBAD database, operated at CDS, Strasbourg, France.
\end{acknowledgements}

\bibliographystyle{aa}
\bibliography{OCMetPaper2,table_mean_paper,table_appendix}

\end{document}